\documentclass[prb,twocolumn,showpacs,preprintnumbers,amsmath,amssymb,superscriptaddress,floatfix]{revtex4-1}
\usepackage{lipsum,babel}
\usepackage{graphicx}
\usepackage{color}
\usepackage[usenames,dvipsnames]{xcolor}
\usepackage{dcolumn}
\usepackage{bm}
\usepackage[latin1]{inputenc}
\usepackage{hyperref}
\usepackage{textcomp}
\usepackage{footnote}
\usepackage{epstopdf}
\usepackage{url}
\usepackage{amsmath}
\usepackage{amssymb}
\usepackage{balance}

\def \MgB2 {MgB$_{2}$ }
\begin{document}
\title{Critical State Theory For The Magnetic Coupling Between Soft Ferromagnetic Materials And Type-II Superconductors}

\newcommand{\orcidauthorA}{0000-0002-1399-5495} % Add \orcidB{} behind the author's name
\newcommand{\orcidauthorB}{0000-0002-6100-1918} % Add \orcidB{} behind the author's name

\author{M. U. Fareed}
\email[Electronic address: ]{muf2@leicester.ac.uk}
\affiliation{Space Park Leicester \& the College of Science and Engineering, University of Leicester, University Rd,  Leicester LE1 7RH, United Kingdom}

\author{H. S. Ruiz}
\email[Electronic address: ]{hsrr1@leicester.ac.uk}
\affiliation{Space Park Leicester \& the College of Science and Engineering, University of Leicester, University Rd,  Leicester LE1 7RH, United Kingdom}

%\date{\today}

%%%%%%%%%%%%%%%%%%%
%%%%%    ABSTRACT        %%%%%%
%%%%%%%%%%%%%%%%%%%

%
\begin{abstract}

Improving our understanding of the physical coupling between type-II superconductors (SC) and soft ferromagnetic materials (SFM), is root for progressing onto the application of SC-SFM metastructures in scenarios such as magnetic cloaking, magnetic shielding, and power transmission systems. However, in the latter some intriguing and yet unexplained phenomena occurred, such as a noticeable rise in the SC energy losses, and a local but not isotropic deformation of its magnetic flux density. These phenomena, which are in apparent contradiction with the most fundamental theory of electromagnetism for superconductivity, i.e., the critical state theory (CST), have remained unexplained for about 20 years, given place to the acceptance of the controversial and yet paradigmatic existence of the so-called overcritical current densities. Therefore, aimed to resolve these long-standing problems, we extended the CST by incorporating a semi-analytical model for cylindrical monocore SC-SFM heterostructures, setting the standards for its validation with a variational approach of multipole functionals for the magnetic coupling between Sc and SFM materials. It is accompanied by a comprehensive numerical study for SFM sheaths of arbitrary dimensions and magnetic relative permeabilities $\mu_{r}$, ranging from $\mu_{r}=5$ (NiZn ferrites) to $\mu_{r}=350000$ (pure Iron), showing how the AC-losses of the SC-SFM metastructure radically changes as a function of the SC and the SFM radius for $\mu_{r} \geq 100$. Our numerical technique and simulations revealed also a good qualitative agreement with the magneto optical imaging observations that were questioning the CST validness, proving therefore that the reported phenomena for self-field SC-SFM heterostructures can be understood without including the ansatz of overcritical currents.

\end{abstract}
\maketitle

%%%%%%%%%%%%%%%%%%%%%%%%%%%%%%%%%%%%%%%%%%%%%%%%%%%%%%%%%%%%%%%%%%%%%%%%%%%%%%%
%% SECTION 1
%%%%%%%%%%%%%%%%%%%%%%%%%%%%%%%%%%%%%%%%%%%%%%%%%%%%%%%%%%%%%%%%%%%%%%%%%%%%%%%

\section{Introduction}\label{Sec.1}

Due to the novel phenomena and applications that can be envisaged by the use of metamaterials, in recent years the developing of superconducting-ferromagnetic metastructures has been the object of considerable attention~\cite{Glowacki2009,Sanchez2011,Navau2014,Solovyov2015,Genenko2015,Gomory2012,Gomory2015,Souc2016,Badia2016,Ruiz2019IEEE,Ruiz2018SciRep}. Particular focus has been played onto the magnetization and demagnetization properties of type-II superconductors (SC) surrounded or in the near proximity of a soft ferromagnetic material (SFM)~\cite{Ruiz2018SciRep,Genenko2004APL}, the study of magnetic cloaking heterostructures~\cite{Souc2016,Genenko2015,Solovyov2015,Gomory2015,Gomory2012}, and their magnetic shielding properties~\cite{Pang1981,Kirchmayr1996,Itoh1996,Lousberg2010,Lousberg2011,Horvat2008,Horvat_2005_SUST,Kovac2006,Gozzelino2013,Gozzelino2017}. Nevertheless, the influence of the physical coupling between the macroscopic electromagnetic properties of the SC and the SFM onto the hysteresis losses of these heterostructures is yet to be understood.

Several semi-analytic approaches for the magnetic shielding properties of SC materials surrounded by soft high-permeability magnets have been already proposed for some configurations, including infinitely thin superconducting strips
~\cite{Genenko2000PRB,Genenko2000,Genenko2002,Genenko2002PRB,Genenko2003,Genenko2006,Genenko2007,Genenko2011,Genenko2012},  cylindrical tubes~\cite{Genenko2004,Genenko2005,Genenko2015}, and finite rounded filaments~\cite{Genenko2004APL,Genenko2005PRB,Genenko2006JoP}. However, a direct involvement of the inductive coupling elements between the profiles of current density in SC wires and finite SFM sheaths is still to be achieved, such that the understanding of the actual physical mechanism that couple their macroscopic magnetic features is not hindered. In fact, a theoretical explanation for the increment of the AC losses in monocore SC-SFM heterostructures at self-field conditions, i.e., under applied transport current but no external magnetic field, has not been reached for even the simplest configuration of a SC-SFM wire of cylindrical cross section, a problem that has remained open for about two decades~\cite{Eckelmann_1998,Huang_1998,Majoros2000}. 

Similarly, by Magneto Optical Imaging (MOI) techniques and the indirect calculation of the SC critical current density by magnetization measurements~\cite{Pan2003,Pan2004a}, an intriguing and yet unexplained modification of the magnetic flux distribution within the SC core of Iron sheathed MgB$_{2}$ monocore wires has been observed, without introducing additional pinning centres. In this regard, akin to the concept of overcritical currents originally introduced for infinitely thin strips in the proximity of a SFM~\cite{Genenko2000}, it was initially thought that this local deformation in the magnetic flux was caused by the occurrence of  overcritical current densities at the so-called flux-free regions~\cite{Roussel2006}. In other words, where the SC can apparently develop regions where the Bean's law of the critical state theory (CST), $J \leq J_{c0}$, with $J_{c0}$ the critical current density at self-field conditions, is violated without destroying the SC state.  

Nevertheless, although it is true that the shielding properties of the SFM can enhance the critical current density of MgB$_{2}$-Fe wires~\cite{Horvat_2002_APL}, as the MgB$_{2}$ is known to show a magnetic field dependence on the critical current density~\cite{Horvat_2005_SUST}, $J_{c}(H)$, these overcritical current densities have not been observed neither by MO techniques~\cite{Roussel2006,Roussel2007,Wells2011} nor by the direct measurement of $J_{c0}$ by electric transport measurements~\cite{Horvat2008}, therefore precluding their existence (at least) in this geometry. However, it is precisely for this geometry where a certain amount of magnetic field has been observed in regions where no transport current is expected to flow, at least under the classical conception of the CST regime for a bare SC at self-field conditions. Also, a significant rise and drop of the local magnetic field within the SC core near the surface of the SFM sheath has been observed~\cite{Roussel2006}, being both of these features in apparent disagreement with the CST, despite its largely recognized success for all known type-II superconductors~\cite{Ruiz2009bPRB,Ruiz2011SUST,Ruiz2012APL,Ruiz2013IEEE,Ruiz2013JAP,Ruiz2018aIEEE,Ruiz2018bIEEE,Ruiz2018SUST,Ruiz2019IEEE,Ruiz2019MDPI}.

%%%%%%%%%%%%%%%%%%%%%%%%%%%%%%%%%%%%%%%%%%%%%%%%%%%%%%%%%%%%%%%%%%%%%%%%%%%%%%%
%% FIGURE 1
%%%%%%%%%%%%%%%%%%%%%%%%%%%%%%%%%%%%%%%%%%%%%%%%%%%%%%%%%%%%%%%%%%%%%%%%%%%%%%%
%%%%%%%%%%%%%%%%%%%
\begin{figure}[t]
\begin{center}
{\includegraphics[width=0.47\textwidth]{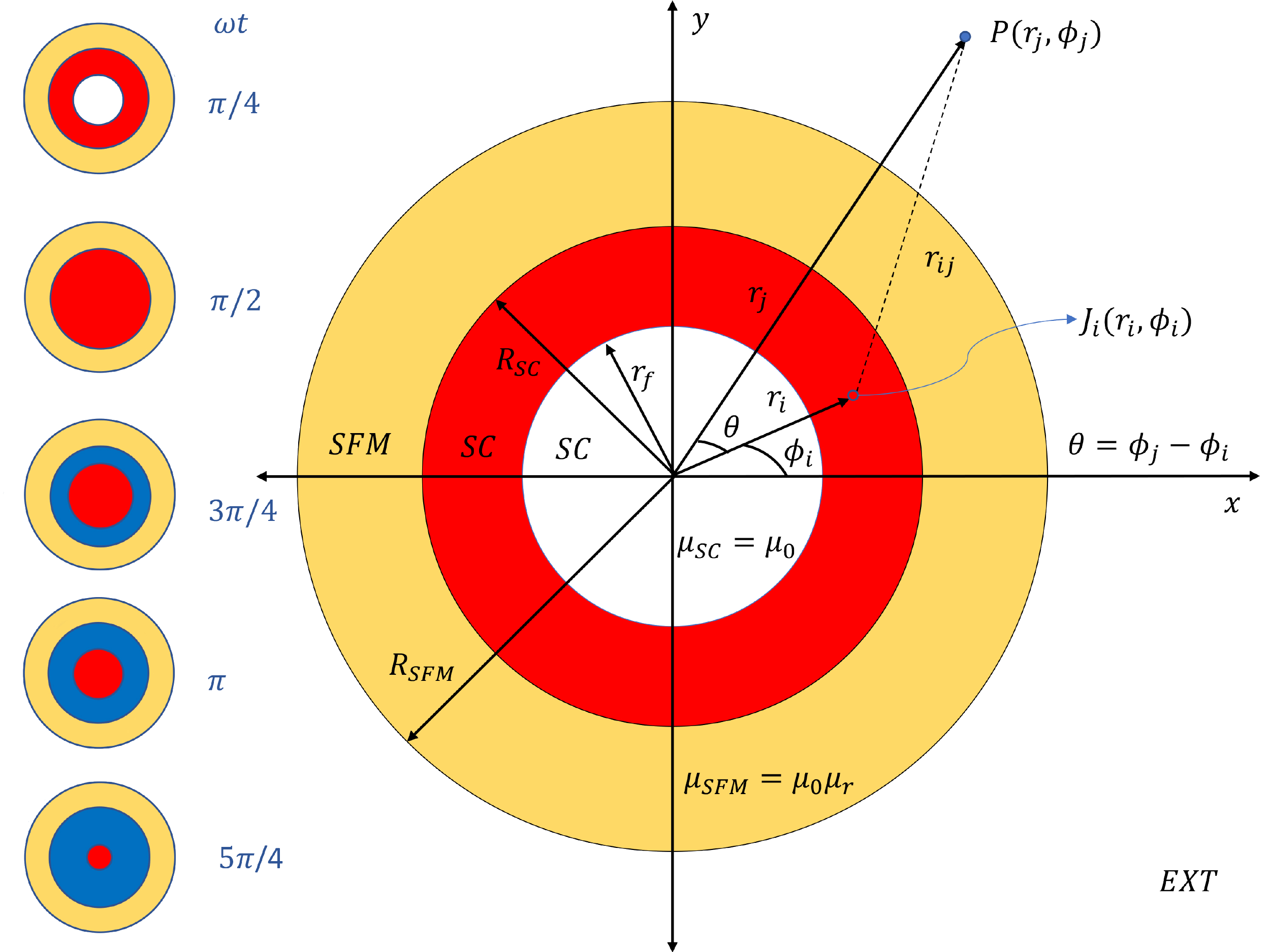}}
\caption{\label{Fig_1} Pictorial representation of the analysed Superconducting (SC) - Soft-Ferromagnetic (SFM) metastructure. The main plot shows the distribution of current density, $\textbf{J}$, in the SC under self-field conditions for an applied transport current $I_{tr}=I_{c}\sin{(\omega t)}$, with $\omega t=\pi/4$ (red shadowed area), and the relative coordinates for a finite-element  $J_{i}(r_{i},\phi_{i})$ as reference for Eqs.~\ref{Eq_5}-\ref{Eq_20}. The left-plots show for illustration the time dynamics of the superconducting current density $\textbf{J}$ within the SC-SFM metastructure. Red and blue areas correspond to distributions with $\textbf{J}_{i}=\pm J_{c}\hat{u}_{z}$, respectively. White areas correspond to regions with no $\textbf{J}$, and the yellow area corresponds to the SFM sheath where no $I_{tr}$ is to flow.}
\end{center}
\end{figure}
%%%%%%%%%%%%%%%%%%%

The aforementioned problems have been somehow ignored, not only due to the engineering prospects of reducing the AC losses in multifilamentary superconductors by the magnetic screening effect of the SFM coatings, another problem that is still to be solved~\cite{Kovac_2003_SUST,Horvat_2005_SUST,Young_2007_IEEE,Majoros_2009_IEEE,Nikulshin_2018_IEEE,Nikulshin_2019_SUST,Xi_2019_IEEE}, but more importantly, it due to the intrinsic difficulty added by the uncertainty on the physical mechanism that couples the electromagnetic properties of SCs and SFMs at a local level (i.e., inside both materials but within a macroscopical approach). Even in the most ideal of the cases, a perfectly cylindrical type-II SC wire of infinite length obeying the general CST~\cite{Ruiz2009bPRB}, i.e., a case where a fully analytic solution for the time dynamics of the flux-front profiles exists~\cite{Gurevich1997}, it results apparently impossible to determine the cause of the increment in the AC-losses for a SC embedded within a closed SFM sheath. This is because the current distribution inside the SC does not change, at least within the quasi-steady low frequency regime where the standard CST applies (below radio frequencies of $\sim$20~kHz)~\cite{Fiorillo2011,Coffey_1992_PRL,Coffey_1992,CoffeyClem_1992} . Thence, it is worth mentioning that for the case of transport current applications in conventional power systems ($\sim$50-60~Hz), the standard CST allows to capture all the electromagnetism of the rounded SC wire under self-field conditions (See Fig.~\ref{Fig_1}), by means of perfectly circular (radial) distributions of uniform current density, for which a fully analytic equation for the calculation of the AC-losses can be derived~\cite{Ruiz2013IEEE,Ruiz2013JAP}. Thus, the cause of these difficulties on the understanding of the AC losses for SC-SFM metastructures is threefold, as it will be explained below.

Firstly, most of the theoretical approaches for the calculation of the AC losses on type-II superconductors start from the assumption of an established formula for the electric field inside the SC, also known as the $\textbf{E}-\textbf{J}$ power law or conductive material law for the SC material~\cite{Grilli2014}. This fact, leaves the entire determination of the AC losses to rely solely on the accurate calculation of the local distribution of current density $\textbf{J}$, which is bounded by the CST law, $\textbf{J }\leq\textbf{ J}_{c}$. However, the distribution and intensity of the current density inside a SC sheathed with a SFM is essentially unaltered, what therefore precludes the idea of getting an increment in the AC losses of the SC. Secondly, the fact that the distribution of magnetic field outside a cylindrical SC-SFM heterostructure under self-field conditions is basically the same than the one of the non-sheathed SC cylindrical wire~\cite{Pan2003,Pan2004a}, makes no reason to think that under this geometry the electric or magnetic field patterns inside the SC material changes, with no apparent change in its critical current density by the influence of the SFM. This actually occurs regardless of the magnetic permeability of the SFM~\cite{Genenko2000PRB}, what makes really difficult to understand what could be the possible cause for an increment on the AC-losses when the SFM is not the source of these hysteretic losses. Thirdly, finite element methods based on the solution of the system  of partial differential equations (PDE) established by James Clerk Maxwell, which commonly solve a global PDE system including the surroundings of the SC-SFM heterostructure, either for the magnetic field \textbf{H}, or the magnetic vector potential \textbf{A} (amongst other PDE models)~\cite{Grilli2016}, are doomed to find the same local solution at the SC domain regardless whether the SC is sheathed by a SFM or not. This is because the attained numerical solution still represents the simplest and mathematically valid response for the SC, which simply neglects any possible magnetostatic coupling between this and the SFM, unless it would have been explicitly included into the numerical formulation.

Therefore, inspired by the pioneering research on circular magnets for high-energy particle accelerators at CERN~\cite{Devred1998,Devred1999a,Devred1999b,Devred2005} and the general CST by Bad\'{\i}a, L\'{o}pez and Ruiz~\cite{Ruiz2009bPRB}, in this paper we included a multipole expansion onto the integral formulation of the CST for type-II SC rounded wires~\cite{Ruiz2011SUST,Ruiz2012APL,Ruiz2013IEEE,Ruiz2013JAP,Ruiz2018aIEEE,Ruiz2018bIEEE,Ruiz2018SUST,Ruiz2019IEEE}, allowing a direct inclusion of the magnetostatic coupling between the SC and a rounded SFM sheath (Sec.~\ref{Sec.2}). In this way, we disclose the electromagnetic behavior of the current density and magnetic field resulting by the coupling of the SC and the SFM in Sec.~\ref{Sec.3}, explaining by semi-analytical and numerical methods the actual causes behind the increment of the AC losses in a SC-SFM cylindrical metastructure in Sec.~\ref{Sec.4}. It has allowed us to conclude how the AC losses of rounded SC-SFM metastructures at self-field conditions can be affected by the amplitude of the transport current and the magnitude of the relative magnetic permeability of the SFM as part of the main conclusions of this study (Sec.~\ref{Sec.5}).

%%%%%%%%%%%%%%%%%%%%%%%%%%%%%%%%%%%%%%%%%%%%%%%%%%%%%%%%%%%%%%%%%%%%%%%%%%%%%%%
%% SECTION 2
%%%%%%%%%%%%%%%%%%%%%%%%%%%%%%%%%%%%%%%%%%%%%%%%%%%%%%%%%%%%%%%%%%%%%%%%%%%%%%%

\section{Multipole Expansion of the CST in Rounded SC-SFM Heterostructures}\label{Sec.2}

The fundamentals of the variational theory for the electromagnetic modeling of type-II superconductors~\cite{Badia2001,Badia2002}, and therefore SC-SFM heterostructures, roots into the application of the optimal control theory for the minimization of the electromagnetic Lagrangian, $\rm{Min}\{\cal{L}\}\equiv\rm{Max}\{\textbf{J}\cdot\textbf{p}\}$,  which is equivalent to the maximum projection rule of the power density, being the electromagnetic Lagrangian multiplier defined as $\textbf{p}=-\Delta\textbf{A}=\textbf{E}\Delta{t}$ for arbitrary variations of the magnetic field $\Delta\textbf{B}=-\nabla\times\textbf{p}$~\cite{Ruiz2009bPRB}. 

Thus, based upon this framework, a small linear path-step between two successive profiles of magnetic field, $\Delta\textbf{B}=\textbf{B}_{n+1}-\textbf{B}_{n}$, can fulfill Amp\`{e}re's law, $\nabla\times\textbf{B}_{n}=\mu_{0}\textbf{J}_{n}$, as well as the continuity conditions $\nabla\cdot{\textbf{B}_{n}}=0$ and $\nabla\cdot{\textbf{J}_{n}}=0$, by imposing the minimization of the step variation for the magnetic field profile integral across the whole $\Re^{3}$-space,
\begin{eqnarray}\label{Eq_1}
{\cal F}
[{\bf B} ( \cdot )] = {\rm Min} \int _{ \Re ^{3}}  \frac{1}{2} \vert {\Delta \bf B} \vert^{2}
\, ,
\end{eqnarray}%
where the SC domain, $\Omega_{SC}$, is conditioned either to the $\textbf{E}-\textbf{J}$ material law, or to the inequality constraint $\textbf{J}\leq\textbf{J}_{c}$ within the SC critical state model. Likewise, the minimization functional must be solved within the excitation dynamics, which in the case of an applied transport current, $I_{tr}$, corresponds to satisfying the condition, 
\begin{eqnarray}\label{Eq_2}
\int _{SC} {\bf J} \cdot  \hat{\bf n}~d\Omega = {\bf I}_{tr}
\, . 
\end{eqnarray}

Then, for a 2D system of $\Omega$-domains (SC, SFM, EXT) as the one considered in Fig.~\ref{Fig_1}, where the elements of current density can only flow along the $z-$axis within the SC domain, i.e., where the dynamics of flux front profiles is restricted to the $x-y$ plane, the minimization functional can be rewritten as, 
\begin{eqnarray}\label{Eq_3}
{\cal F}[{\bf A} ( \cdot )] = {\rm Min} \int _{{\Re}^{2}} [ \Delta {\bf A}_{z} \cdot {\bf J} + \nabla\Phi\cdot{\textbf{J}} \Delta t ]
\, .
\end{eqnarray}

In this case, the gradient of the scalar electric potential $\nabla\Phi=C_{t}\hat{u}_{z}$ is different to zero only if $I_{tr}\neq 0$, with the electric field and magnetic vector potential directed along the $z-$axis, $\textbf{E}_{z}=-\partial_{t}\textbf{A}+C_{t}$, being $C_{t}$ the integration constant. Thus, as the AC-losses of the system are determined by the integral of the instantaneous power density losses across the material domains, over a hysteresis cycle of the transport current excitation of frequency $\omega$, i.e., 
\begin{eqnarray}\label{Eq_4}
L=\omega \oint_{t}\int_{\Omega} \textbf{E}\cdot \textbf{J}~d\Omega ~ d{t}
\, ,
\end{eqnarray}
then, the problem simply reduces to determine the magnetic vector potential across the different material domains with current density. Therefore, in a first approach for a cylindrical wire of radius $R_{SC}$, it seems impossible to predict a rise in the AC-losses of SC-SFM metastructures under self field conditions, as not only induced magnetic losses by the non-hysteretic SFM can be neglected~\cite{Genenko2004APL,Genenko2006JoP}, but also no current sharing is to be seen between the SC and the SFM material~\cite{Vlasko2015}. 

Thus, with no magnetization losses neither current profiles within the SFM, and exactly the same distribution of current density inside the SC, the boundary of the flux-front profile for a cylindrical SC wire can be determined by exact analytical methods~\cite{Gurevich1997}, following the area enclosed between the surface of the SC and a circumference of radius,
\begin{eqnarray}\label{Eq_5}
r_{f}=R_{SC}\sqrt{1-\frac{I_{tr}}{I_{c}}}~\, , 
\end{eqnarray}
being $r_{f}$ the inner boundary of the flux front profile as shown in Fig.\ref{Fig_1}. Then, any possible change in the losses of the system will be restricted to the definition of the electric field invoked into the SC domain. 

The above result allows to immediately identify why conventional PDE solvers such as COMSOL Multiphysics cannot predict the increment in the AC losses in the SC domain, as the material law that governs the physics of the macroscopic magnetic behavior of a SC, is directly entered by the empirical ansatz known as the $\textbf{E}-\textbf{J}$ power law. This so-called law, although very acknowledged for reproducing the electromagnetic behavior of practical type-II superconductors and its applications, also forces the electric field to be a known function which is primarily measured at the condition of self-field critical current density. Therefore, it does not take into consideration any intrinsic variance within the magnetic vector potential $(\textbf{A})$, nor any possible contribution by other materials such as a SFM. 

However, within the integral formulation of Eq.~\ref{Eq_1}, and consequently for Eq.~\ref{Eq_2}, the electric field is seldom calculated by the use of empirical material laws for the SC state, but instead from the well established Bean's theorem for the CST~\cite{Bean1962}, and by the knowledge of an analytical function for the magnetic vector potential along the ${\Re}^{3}$ space~\cite{Badia2001}. This in turn can be transformed into a function of finite elements of current density $\textbf{J}_{i}$, that multiplied by their inductance matrices, i.e., by terms which do not depend on any physical variable but on the position of the elements of current, $\textbf{r}_{i}$ (see Fig.~\ref{Fig_1}), leads to a reduction on the size and dimensionality of the minimization integral from the whole ${\Re}^{3}$-space~\cite{Ruiz2009bPRB,Ruiz2012APL,Ruiz2011PRB}, to just the volume or area of the SC domain, $\Omega_{SC}$. Therefore, if for the 2D geometry shown in Fig.~\ref{Fig_1}, it is assumed that the elements of current density are to appear only inside the SC domain, then, in the absence of the SFM, these elements can be treated as infinitely long and thin wires with the vector potential for the self and mutual inductances defined by, 
\begin{eqnarray}\label{Eq_6}
\textbf{A}_{i}(\textbf{r}_{i})=(\mu_{0}/4\pi)\pi\textbf{J}_{i} ~\, , 
\end{eqnarray}
\begin{eqnarray}\label{Eq_7}
\textbf{A}_{ij}(\textbf{r}_{j})=-(\mu_{0}/4\pi)\ln{(r_{ij}^{2})} \textbf{J}_{i} \,\,\, \forall \,\,\, r_{ij}\neq 0 \, , 
\end{eqnarray}
being $r_{ij}$ the distance between two lines of current each at the positions $r_{i}$ and $r_{j}$.

Then, in order to formulate the magnetic vector potential in the case of a SC-SFM metastructure, the starting point is to define the distance $r_{ij}$ in the complex or $s-plane$ as $r_{ij}=r_{i}-s$, with $s=r_{j}e^{i\theta}$ (see Fig.~\ref{Fig_1}), such that the real part of the vector potential, $Re\{\textbf{A}\}=A_{z}$, is defined by the vector potential created by a line of current $\textbf{J}_{i}$ at any position $\textbf{r}_{j}\neq\textbf{r}_{i}$ as
\begin{eqnarray}\label{Eq_8}
\textbf{A}_{ij}(\textbf{r}_{j})=-\frac{\mu_{0}}{2\pi}\textbf{J}_{i}\ln{\left(\kappa~r_{ij} \right)} \,\,\, \forall \,\,\, r_{j}\in \kappa
~\, .
\end{eqnarray}
Here, the index $\kappa=\pm 1$ separates the space into two conditions, one for $0<r_{j}<r_{i}$ when $\kappa=1$, i.e., a condition that is commonly found at beam optics computations in the case of accelerator magnets~\cite{Devred1998}, and the other, the condition for $r_{j}>r_{i}$ with $\kappa=-1$, which results useful for magnetic computations~\cite{Devred1999a}. Then, the key instrument is to expand the function $\ln{(r_{ij})}=\ln{(r_{i})}+\ln{(1-s/r_{i})}$ into a Taylor's series, that with the help of De Moivre's Formula allows to rewrite Eq.~\ref{Eq_8} as:
\begin{eqnarray}\label{Eq_9}
{\bf A}_{i,j} = - \frac{\mu_{0}}{2\pi} {\bf J}_{i} \left[ \ln{(r_{k})}- \sum_{n=1}^{\infty} \frac{1}{n} \left(\frac{r_{j}}{r_{i}}\right)^{\kappa n} cos(n\theta)\right] \, ,
\end{eqnarray}
with $\kappa=1$ for $r_{k}=r_{i}$, and $\kappa=-1$ for $r_{k}=r_{j}$, where it is to be noted that the vector potential is continuous at $r_{j}=r_{i}$. Therefore, when considering the SFM medium, the problem can be solved by means of the Laplace's equation $\square^{2} {\bf A} = \nabla^{2} {\bf A} - \partial_{t}^{2}{\textbf{A}}/c^{2}=-4\pi c^{-1}\mu_{r} {\bf J}$, which in the magneto quasi-steady approach introduced in Ref.~\cite{Ruiz2009bPRB}, i.e., with $\partial_{t}^{2}{\textbf{A}}=0$, it can be simplified to $c\nabla^{2} {\bf A}=0$, as no current sharing can be assumed between the SC and the SFM. Thus, in the case of a SC-SFM metastructure as the one shown in Fig.~\ref{Fig_1}, this equation can be solved in cylindrical coordinates by the method of separation of variables, such that its solution can be expressed as $A_{c,m}=R(r)\Theta(\theta)$, with $\Theta(\theta)$ a $2\pi$ periodic function of $\theta$ and, the Laplace' equation simplified to 
\begin{eqnarray}\label{Eq_10}
\frac{r}{R}\frac{\partial}{\partial r}\left(r \frac{\partial R}{\partial r} \right) = -\frac{1}{\Theta}\frac{\partial^{2}\Theta}{\partial\theta^{2}} = C\, ,
\end{eqnarray}
with $C$ a real constant that does not depend on $r$ nor $\theta$. 

Consequently, the most general solution to Eq.~\ref{Eq_10} is a linear superposition for all possible solutions~\cite{Devred2005}, either with $C=0$, $C>0$, or $C<0$, resulting in the general definition for the vector potential in the absence of current density for a coupled medium $m$,
\begin{eqnarray}\label{Eq_11}
\textbf{A}_{c,m} = && E_{0,m} C_{0,m}+D_{0,m}\ln{(r)} \nonumber + ... \\ &&  \sum_{n=1}^{\infty} E_{n,m} cos (n \theta)\left(C_{n,m} r^{n} + D_{n,m} r^{-n}\right) 
\, ,
\end{eqnarray}
with the unknown media-dependent parameters $C_{0,m}$, $C_{n,m}$, $D_{0,m}$, $D_{n,m}$, $E_{0,m}$, and $E_{n,m}$, all being real integration constants which can be determined by superimposing both, the vector potential created by the mere existence of a media, i.e., Eq.~\ref{Eq_11}, with the one created by the existence of a line of current, i.e., Eq.~\ref{Eq_9} and, by further imposing adequate boundary conditions at the interfaces between the different media. Thus, for the SC-SFM metastructure shown in Fig.~\ref{Fig_1}, the magnetic vector potential created by a line of current $J_{i}$ located at $r_{i}$ over a point in the space $r_{j}\neq r_{i}$, i.e., $A_{z,m}(r_{j})=A_{i,j}(r_{k})+A_{m}(r_{j})$,  must be defined within four different regions of the space, two of these within the SC domain for the conditions $\kappa=1$ (i.e., $0<r_{j}<r_{i}$) and $\kappa=-1$ (i.e., $r_{i}<r_{j}<R_{SC}$), and the other two defining the space occupied by the SFM layer $(R_{SC}<r_{j}<R_{SFM})$, and the outer domain (EXT) defined by the condition $R_{SFM}<r_{j}$ (see Fig.~\ref{Fig_1}). 

Then, in order to get an unequivocal physical solution, this system of equations must satisfy continuity boundary conditions at $r_{j}=R_{SC}$ and $r_{j}=R_{SFM}$, i.e., at the interfaces between the two different mediums. In other words, an additional set of equations both for the magnetic vector potential $A_{z}$, and the magnetic field vector $\textbf{B}=\nabla\times\textbf{A}_{z}$, is established by having into consideration the conditions $A_{z,SC}(R_{SC})=A_{z,SFM}(R_{SC})$, and $A_{z,SFM}(R_{SFM})=A_{z,EXT}(R_{SFM})$. In this way,  the condition of non divergence of the magnetic field is preserved by satisfying the conditions $\partial_{r}A_{z,SC}=\mu_{r}^{-1}\partial_{r}A_{z,SFM}$, and $\partial_{r}A_{z,EXT}=\mu_{r}^{-1}\partial_{r}A_{z,SFM}$, for $r_{j}=R_{SC}$ and $r_{j}=R_{SFM}$, respectively. This creates a set of minimum six equations at each one of the interfaces between the different media, from which we can determine the set of six media-dependent constants $C_{0,m}$, $C_{n,m}$, $D_{0,m}$, $D_{n,m}$, $E_{0,m}$, and $E_{n,m}$. Thus, after some algebra, it is possible to demonstrate that the total vector potential for the different media (SC, SFM, and EXT) in Fig.~\ref{Fig_1}, can be written as:
\begin{eqnarray}\label{Eq_12}
&&\textbf{A}_{SC}(\textbf{r}_{j}\leq R_{SC})=\textbf{A}_{j}(\textbf{r}_{j})+\sum_{i \neq j}\textbf{A}_{ij}(\textbf{r}_{j})+\textbf{A}_{c,SC}(\textbf{r}_{j})  \, , \nonumber \\
&&\textbf{A}_{SFM}(R_{SC}<\textbf{r}_{j}\leq R_{SFM})=\sum_{i}\textbf{A}_{ij}(\textbf{r}_{j})+\textbf{A}_{c,SFM}(\textbf{r}_{j}) \, , \nonumber \\
&&\textbf{A}_{EXT}(R_{SFM}<\textbf{r}_{j})=\sum_{i}\textbf{A}_{ij}(\textbf{r}_{j})+\textbf{A}_{c,EXT}(\textbf{r}_{j}) \, ,
\end{eqnarray}
with the vector potentials for the coupled media defined for the conditional $\mu_{(\pm)}=\mu_{r}\pm 1$ by,
\begin{widetext}
\begin{eqnarray}\label{Eq_13}
\textbf{A}_{c,SC}=
-\frac{\mu_{0}}{2\pi}\mu_{(-)}\textbf{J}_{i}
\left[
\mu_{(+)}\sum_{n=1}^{\infty} \frac{\bar{R}_{\mu 1}}{n} 
\left(\frac{r_{i}r_{j}}{R_{SC}^{2}}\right)^{n} 
\cos(n\phi_{j}) 
\right]
\, ,
\end{eqnarray}
\begin{eqnarray}\label{Eq_14}
\textbf{A}_{c,SFM}=
\frac{\mu_{0}\mu_{(-)}}{2\pi} \textbf{J}_{i }
\left[ 
\ln\left(\frac{R_{SC}}{r_{j}} \right) 
- \sum_{n=1}^{\infty} \frac{\bar{R}_{\mu 2-}}{n} \left(\frac{r_{i}}{r_{j}}\right)^{n}
\cos(n\phi_{j}) 
\right]
\, ,
\end{eqnarray}
and,
\begin{eqnarray}\label{Eq_15}
\textbf{A}_{c,EXT}=
\frac{\mu_{0}}{2\pi}\mu_{(-)} \textbf{J}_{i }
\left[ 
\ln\left(\frac{R_{SC}}{R_{SFM}} \right)   
+ \mu_{(-)}\sum_{n=1}^{\infty} \frac{\bar{R}_{\mu 1}}{n} \left(\frac{r_{i}}{r_{j}}\right)^{n} 
\cos(n\phi_{j}) 
\right]
\, ,
\end{eqnarray}
\end{widetext}
with,
\begin{eqnarray}\label{Eq_16}
\bar{R}_{\mu 1}=\frac{R_{SFM}^{2n}-R_{SC}^{2n}}{\mu_{(-)}^{2} R_{SC}^{2n} - \mu_{(+)}^{2} R_{SFM}^{2n}} \, ,
\end{eqnarray}
and
\begin{eqnarray}\label{Eq_17} 
\bar{R}_{\mu 2\pm}=\frac{\mu_{(+)}R_{SFM}^{2n} \pm 2\mu_{r} r_{j}^{2n} + \mu_{(-)} R_{SC}^{2n}}{\mu_{(-)}^{2} R_{SC}^{2n} - \mu_{(+)}^{2} R_{SFM}^{2n}} 
\, ,
\end{eqnarray}
such that if the magnetic properties of the SFM are removed, i.e., if its magnetic permeability takes the value of the relative magnetic permeability of vacuum, $\mu_{r}=1$, then all the coupling contributions in Eq.~\ref{Eq_12} disappear, returning to the classical problem where the distribution of current into a superconductor can be calculated by the simple knowledge of the self and mutual inductance matrices for finite elements of critical current density~\cite{Ruiz2012APL}. On the other hand, for understanding the coupling elements between the SC and the SFM, we have introduced the non-dimensional factors $\bar{R}_{\mu 1}$ and $\bar{R}_{\mu 2\pm}$, where it is to be noticed that $\bar{R}_{\mu 2\pm}$ is not a constant but a function of the element coordinate $r_{j}$. Also, it is to be noticed that these contributions are a response of the SFM to the lines of current density $\textbf{J}_{i}$ inside the SC, which are calculated through the minimization functional shown in Eq.~\ref{Eq_2}. 

Therefore, within the integral formulation in Eq.~\ref{Eq_3} and the vector potentials obtained in Eqs.~\ref{Eq_12}-\ref{Eq_15}, the system is reduced to the calculation of the profiles of current density inside the SC only, providing a tremendous advantage against any other computational method. This is because not only the coupling between the SC and SFM are explicitly included, but also because the infinite $\Re^{2}$-space has been reduced to just the area occupied by the SC domain, in contrast with the use of the whole $\Re^{3}$-space in the case of the differential formulations. Additionally, although the coupling terms depend on the reach of the n-index for the introduced Taylor's series, it is to be noticed that the arguments of these summations are purely geometrical, therefore defining the multipole coefficients for the coupling inductance matrices between the SC and the SFM materials. In this sense, for a given finite element mesh, these matrices can be univocally calculated outside of the minimization process, with the resulting matrix for the positions $r_{i}$ and $r_{j}$ being  stored as a matrix of constant parameters into the minimization algorithm, thence, substantially reducing its computing time. Thus, the only limitation of this method lies in the computational limits and numerical precision for the calculation of the multipole coefficients for the coupled-media vector potentials, which refers to the smallest and the largest positive normalized floating-point number in IEEE double precision, i.e., $2^{-1022}$ and $(2-2^{-52})*2^{1023}$, respectively. 

Consequently, if the SC and SFM radii are written in normalized units such that, $R_{SC}=1$, and $R_{SFM}=1.5 R_{SC}$, for the effects of the minimization process of the functional of interest, ${\cal F}[{\bf A} ( \cdot )]={\cal F}_{SC}[\textbf{A}_{SC} ( \textbf{J}_{i} ),C_{t}]$, the largest n-index that could be considered is $n=\log(1.7977\times 10^{308})/\log(1.5)\simeq 1750$, from which we have found that within a $10^{-8}$ tolerance factor, any n-index greater than $\sim350$ will produce the same results. Then, by knowing the total magnetic vector potential across the whole space, it is possible to numerically determine the distribution of current density $\textbf{J}_{i}$ inside the SC-SFM metastructure for a given time, by solving the minimization functional ${\cal F}[\textbf{A}_{SC} ( \textbf{J}_{i} ), C_{t}]$ subject to: (i) the CST condition $|J_{i}|\leq J_{c}$ and, (ii) the applied transport current constraint $I_{tr}(t)=I_{0}sin(\omega t)$ in Eq.~\ref{Eq_2}, with $I_{0}$ the amplitude of the alternating current (AC) of frequency $\omega$. Likewise, the spatial-constant $C_{t}$ that appears into the minimization functional must be introduced as a time-dependent variable into the numerical procedure~\cite{Ruiz2012APL}, such that the correct value for the electric field and the AC losses can be determined by ensuring that the electric field at the flux free regions satisfy the condition $E_{z}(r_{j}<r_{f})\equiv 0$. 

Only very small increments in the instantaneous magnitude of the electric field inside the SC have been observed by the coupling with the SFM sheath (in the order of $1\times 10^{-3}(\mu_{0}/4\pi)R_{SC}^{2}J_{c}\delta t^{-1}$), such that the local distribution of power density $\textbf{E}\cdot\textbf{J}$ shows not only the same classical behaviour already shown for bare SC wires~\cite{Ruiz2013Book,Ruiz2013IEEE}, but exactly the same distribution of local profiles of critical current density that could be calculated by analytical methods. Thus, although the slight increment in the time-dependent electric field inside the SC ultimately contributes to the increment on the hysteresis losses of the SC-SFM system, it does not provide a very rich physics phenomenology which could reveal the actual impact of the SFM coupling with the SC current. Nevertheless, as the AC losses of SC-SFM metastructures fundamentally depend on the relative magnetic permeability of the SFM, in our attempt to fully answer how the relative magnetic permeability of a SFM sheath affects the AC-losses of a SC wire, we have conducted a large number of simulations (330) including 10 different amplitudes of $I_{0}$, ranging from $0.1I_{c}$ to $I_{c}$. It includes 33 different SFM with relative magnetic permeabilities that range from, $\mu_{r}=5$ for NiZn ferrites~\cite{Williams1992}, up to the very high magnetic permeability measured for the purest Iron, $\mu_{r}=350000$~\cite{Williams1992,DoD1987,Solymar1988}. This comprehensive study has allowed us to unveil the key fingerprint for the most notorious feature of the SC-SFM coupling, which lies in the anomalous distribution of local profiles of magnetic field inside the SC, which is caused by the induced magnetic multipoles created by the interaction between the supercurrents and the SFM sheath.

Thus, either by calculating the distribution of profiles of current density by the minimization functional ${\cal F}[\textbf{A}_{SC} ( \textbf{J}_{i} )]$ or, by directly meshing the distribution of profiles of current density $J_{i}$ within the analytically derived flux front boundary $r_{f}$ (Eq.~\ref{Eq_5}), the magnetic field can be calculated by its general definition $\textbf{B}=\nabla\times\textbf{A}$, which for our 2D cylindrical geometry (see Fig.~\ref{Fig_1}) is reduced to $\textbf{B}=r^{-1}\partial_{\phi}A_{z}\textbf{\^{u}}_{r}-\partial_{r}A_{z}\textbf{\^{u}}_{\phi}$ , where $A_{z}$ is split into the three continuous media $A_{SC}$, $A_{SFM}$, and $A_{EXT}$ at Eq.~\ref{Eq_12}. Therefore, by calculating the corresponding derivatives, we have obtained that at each one of the domains representing these media, the components of the magnetic field can be calculated by the functions, 
\begin{widetext}
\begin{eqnarray}\label{Eq_18}\nonumber
\textbf{B}_{SC}=
\frac{\mu_{0}}{2\pi}J_{i}
\left\lbrace
\begin{array}{ll|}
\left[
\frac{r_{i}}{r_{ij}^{\,2}}\sin(\phi_{i}-\phi_{j})
+ 
\mu_{(-)} \mu_{(+)}\sum_{n=1}^{\infty} \frac{\bar{R}_{\mu 1}}{r_{j}} 
\left(\frac{r_{i}r_{j}}{R_{SC}^{2}}\right)^{n} 
\sin(n\phi_{j}) 
\right] \hat{\textbf{u}}_{r}
\\
\\
\left[
\frac{r_{j}}{r_{ij}^{\,2}}-\frac{r_{i}}{r_{ij}^{\,2}}\cos(\phi_{i}-\phi_{j})
+ 
\mu_{(-)} \mu_{(+)}\sum_{n=1}^{\infty} \frac{\bar{R}_{\mu 1}}{r_{j}} 
\left(\frac{r_{i}r_{j}}{R_{SC}^{2}}\right)^{n} 
\cos(n\phi_{j}) 
\right] \hat{\textbf{u}}_{\phi}
\end{array}
\right\rbrace 
\, ,
\end{eqnarray}
\begin{eqnarray}\label{Eq_19}\nonumber
\textbf{B}_{SFM}=
\frac{\mu_{0}}{2\pi}J_{i}
\left\lbrace
\begin{array}{ll|}
\left[
\frac{r_{i}}{r_{ij}^{\,2}}\sin(\phi_{i}-\phi_{j})
+ 
\mu_{(-)} \sum_{n=1}^{\infty} \frac{\bar{R}_{\mu 2-}}{r_{j}} 
\left(\frac{r_{i}}{r_{j}}\right)^{n} 
\sin(n\phi_{j}) 
\right] \hat{\textbf{u}}_{r}
\\
\\
\left[
\frac{r_{j}}{r_{ij}^{\,2}}-\frac{r_{i}}{r_{ij}^{\,2}}\cos(\phi_{i}-\phi_{j})
+ 
\mu_{(-)} \sum_{n=1}^{\infty} 
\left[
1-\left(\frac{r_{i}}{r_{j}}\right)^{n} \bar{R}_{\mu 2+} 
\right]
\left(\frac{1}{r_{j}}\right) 
\cos(n\phi_{j}) 
\right] \hat{\textbf{u}}_{\phi}
\end{array}
\right\rbrace 
\, ,
\end{eqnarray}
\begin{eqnarray}\label{Eq_20}
\textbf{B}_{EXT}=
\frac{\mu_{0}}{2\pi}J_{i}
\left\lbrace
\begin{array}{ll|}
\left[
\frac{r_{i}}{r_{ij}^{\,2}}\sin(\phi_{i}-\phi_{j})
- 
\mu_{(-)}^{2}\sum_{n=1}^{\infty} \frac{\bar{R}_{\mu 1}}{r_{j}} 
\left(\frac{r_{i}}{r_{j}}\right)^{n} 
\sin(n\phi_{j}) 
\right] \hat{\textbf{u}}_{r}
\\
\\
\left[
\frac{r_{j}}{r_{ij}^{\,2}}-\frac{r_{i}}{r_{ij}^{\,2}}\cos(\phi_{i}-\phi_{j})
+
\mu_{(-)}^{2} \sum_{n=1}^{\infty} \frac{\bar{R}_{\mu 1}}{r_{j}} 
\left(\frac{r_{i}}{r_{j}}\right)^{n} 
\cos(n\phi_{j}) 
\right] \hat{\textbf{u}}_{\phi}
\end{array}
\right\rbrace 
\, .
\end{eqnarray}
\end{widetext}

In this sense, we have arrived to entirely analytical solutions for the magnetic vector potential and the distribution of the magnetic field inside the SC core of a cylindrical SC-SFM metastructure, it subjected to an AC transport current under self-field conditions. This has revealed two important phenomena to be analysed in the following sections. Firstly, demonstrating that the origin of the deformations of the magnetic field inside SC-SFM wires reported by MOI techniques~\cite{Pan2003,Pan2004,Roussel2006,Roussel2007}, results as a direct consequence of the magneto-steady coupling between the SC and the SFM sheath. Secondly, with our extended CST it will be proven that a straightforward explanation of the intriguing increment in the AC-losses of SC-SFM metastructures~\cite{Eckelmann_1998,Huang_1998,Majoros2000}, can be achieved without the ansatz of overcritical currents.

%%%%%%%%%%%%%%%%%%%%%%%%%%%%%%%%%%%%%%%%%%%%%%%%%%%%%%%%%%%%%%%%%%%%%%%%%%%%%%%
%% SECTION 3
%%%%%%%%%%%%%%%%%%%%%%%%%%%%%%%%%%%%%%%%%%%%%%%%%%%%%%%%%%%%%%%%%%%%%%%%%%%%%%%

\section{SC vs SC-SFM Metastructures: Differences on the Current Density and Magnetic Field Profiles}\label{Sec.3}

\begin{figure*}[t]
\begin{center}
{\includegraphics[width=0.95\textwidth]{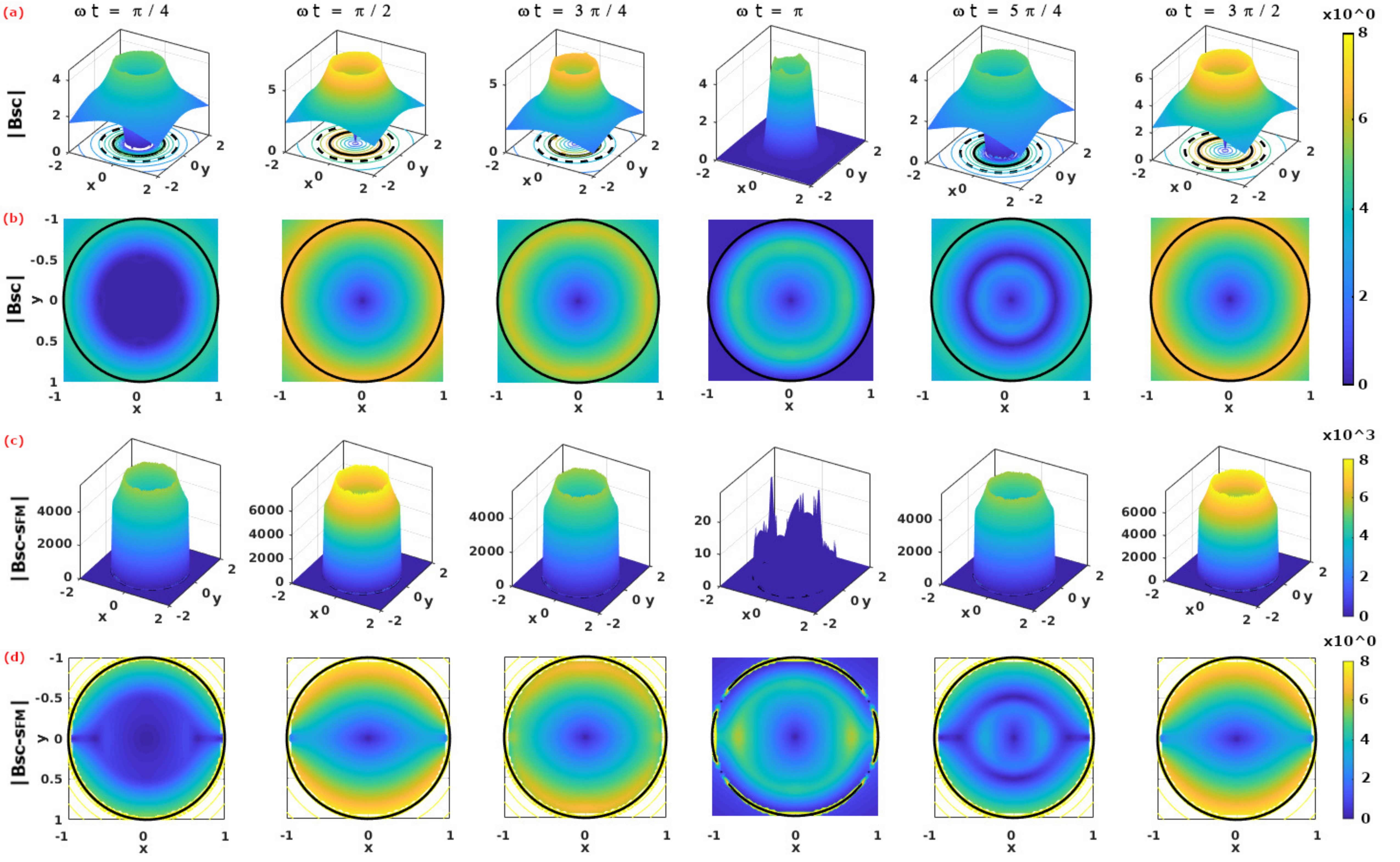}}
\caption{\label{Fig_2} Dynamics of the norm of magnetic flux density $|\textbf{B}|$ in units of $(\mu_{0}/4\pi)J_{c}R_{SC}$ in (a) a SC wire (without SFM sheath) of radius $R_{SC}=1$ (in arbitrary units) whose cross section lies on the plane $xy$ and which is subjected to an applied transport current $I_{tr}=I_{c}\sin(\omega t)$. (b) shows the same distribution of field but in a 2D representation that focus on the local flux dynamics inside the SC, $|\textbf{B}_{SC}|$. Analogously, (c) shows the flux distribution $|\textbf{B}_{SC-SFM}|$ for the SC-SFM metastructure with $R_{SFM}=1.5R_{SC}$ and $\mu_{r}=46$, where the impact of the SFM on the SC can be seen clearer in the 2D representation shown in (d), i.e., the bottom pane of subplots. The time interval between columns is $\Delta t=(\pi/4)\omega^{-1}$, such that the dynamics shown between the second column $(\omega t=3*\pi/2)$ and last column $(\omega t=\pi/2)$ represents the minimum hysteresis period for the calculation of AC Losses, in accordance with the distribution of profiles of current density shown in Fig.~\ref{Fig_1}.
}
\end{center}
\end{figure*}

In Fig.~\ref{Fig_2}, the norm of the magnetic field is shown as a function of the non-dimensional time argument $\omega t$ of the applied transport current, $I _{tr}=I_{0}\sin(\omega t)$, with maximum amplitude, $I_{0}=I_{c}$, illustrating its behaviour during the first ramp of the AC current at $\omega t = \pi/4$ (1st column), as well as during the hysteretic period observed between the peaks $\omega t = 2\pi$ (2nd column) and $\omega t = 3\pi/2$ (6th column), as it suffices for the calculation of the AC-losses when the time integral in Eq.~\ref{Eq_4} is defined between these time-steps and then, multiplied by a factor 2. 

For the sake of comparison, the results presented are shown under two different considerations: (i) the first (top two rows) refers to the case when $\mu_{r}=1$, i.e., in absence of the SFM sheath,  and (ii) the second (bottom two rows) makes reference to the case where the SFM sheath in Fig.~\ref{Fig_1} is defined by a relative radius $R_{SFM}=1.5R_{SC}$ and a magnetic permeability $\mu_{r}=46$, being this a typical magnetic permeability encountered for MgB$_{2}$-Fe wires~\cite{Pan2003,Pan2004,Genenko2004APL,Genenko2005PRB}. Thus, it is to be noticed that as consequence of the magnetic coupling with the SFM sheath, a remarkable deformation of the local density of magnetic flux inside the SC has been found (bottom pane in Fig.~\ref{Fig_2}). This is despite the fact that the distribution of current density still follows the circumferential evolution observed for unsheathed SC wires, i.e., delimited by the flux front analytically derived in Eq.~\ref{Eq_5}, and which has been shown for illustration purposes in Fig.~\ref{Fig_1}. Notoriously, this observation is in remarkable qualitative agreement with the experimental evidences for a characteristic ``elevation'' and ``dip'' of the magnetic flux at self-field conditions, it measured near the SC-SFM interface at the line-angle $r \angle 0$, i.e., at the $x-axis$ from the observer's perspective by Magneto Optical Imaging (MOI) techniques~\cite{Pan2003,Pan2004,Roussel2006,Roussel2007}. Reproducing these experimental results will be the aim of this and the following section, although to understand how the experimental results are reconstructed, it is first necessary to understand the entire dynamics of the electromagnetic quantities along the cross section of the SC-SFM metastructure. 

The MOI observations were initially thought to be in apparent contradiction with the critical state regime, as not only some magnetic field appeared at the so-called \textit{flux-free} regions, i.e, regions where no current density is expected to be flowing, but also, because it does not have a qualitative resemblance with the angular invariant pattern for the magnetic field outside the SC wire, it regardless whether the SC wire has been sheathed or not by a SFM. Thus, this intriguing phenomena which was believed to be caused by some mechanism similar to the overcritical state model in thin SC strips by Genenko et. al.~\cite{Genenko2002,Genenko2000PRB,Genenko2002PRB}, has been motive of a paradigm in superconductivity, as the so-called  overcritical current densities have not been directly observed by electrical measurements in rounded SC-SFM wires~\cite{Roussel2006,Roussel2007,Wells2011,Horvat2008}. Nevertheless, in this paper 
we have demonstrated that the inclusion of the magnetic multipoles created by the physical coupling between the SC and the SFM, are sufficient to reproduce all the macroscopic electromagnetic features of SC-SFM rounded metastructures, without violating the most fundamental principles of the general critical state theory~\cite{Ruiz2009bPRB}.

%%%%%%%%%%%%%%%%%%%%%%%%%%%%%%%%%%%%%%%%%%%%%%%%%%%%%%%%%%%%%%%%%%%%%%%%%%%%%%%
%% FIGURE 2
%%%%%%%%%%%%%%%%%%%%%%%%%%%%%%%%%%%%%%%%%%%%%%%%%%%%%%%%%%%%%%%%%%%%%%%%%%%%%%%
%%%%%%%%%%%%%%%%%%%
%%%%%%%%%%%%%%%%%%%

\begin{figure*}[t]
\begin{center}
{\includegraphics[width=1.0\textwidth]{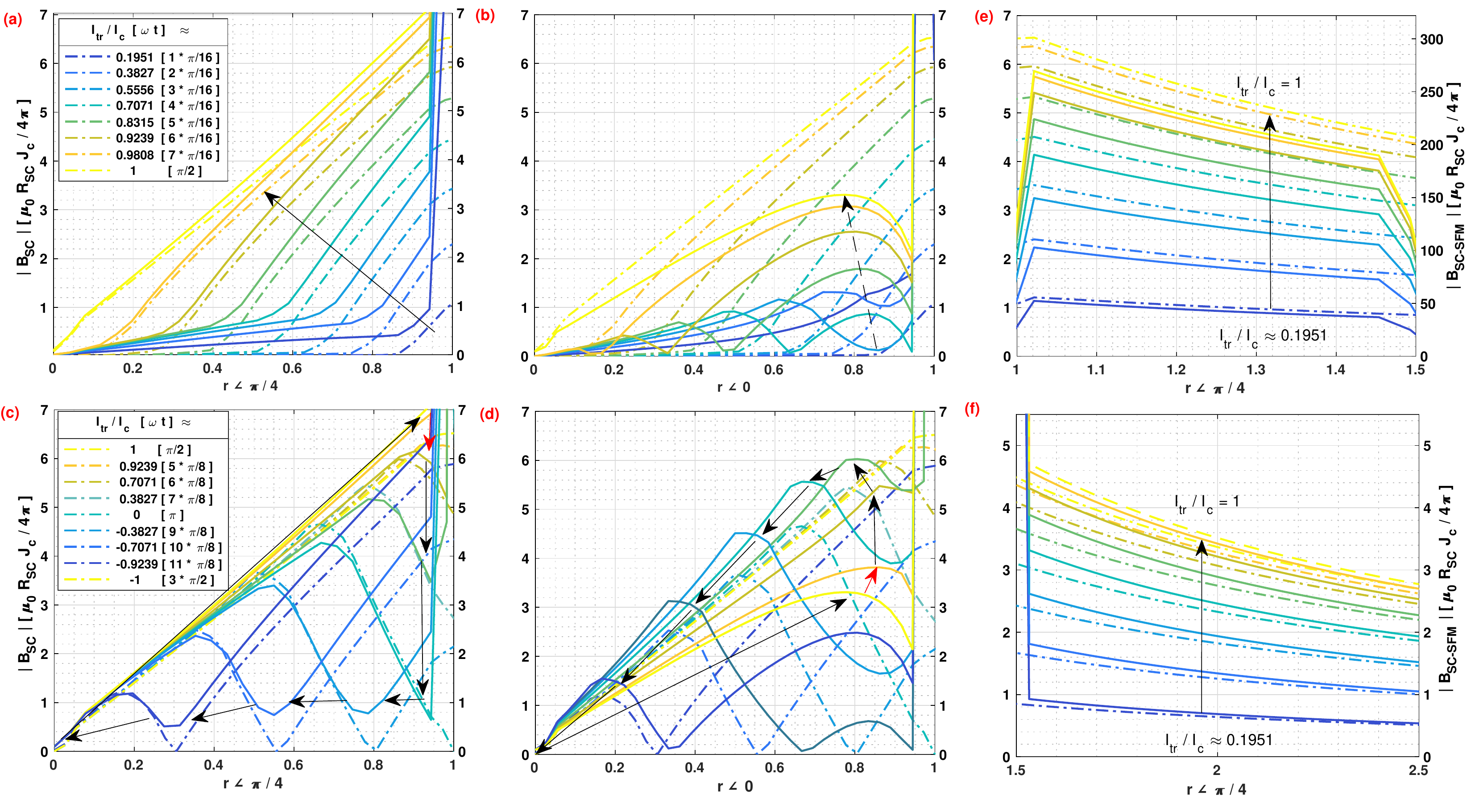}}
\caption{\label{Fig_3}  Dynamics of the norm of magnetic flux density over the radial directions $(r,\pi/4)$ and $(r,0)$ for different magnitudes of the applied AC transport current, $I_{tr}=I_{c}\sin{(\omega t)}$, it measured inside the SC during (a-b) the first ramp of the applied current (top legend-box) and (c-d) the peak-to-peak hysteretic period (bottom legend-box) as described in Figs.~\ref{Fig_1} \& \ref{Fig_2}. The panel of subplots at the right shows the corresponding profiles for the first ramp of current at (e) inside the SFM and (f) outside the SC-SFM wire, respectively. Dashed-dot lines at each subplot refer to the left axes $|\textbf{B}_{SC}|$ showing a classical Bean's behaviour, whilst the solid lines must be read accordingly with the right hand axes $|\textbf{B}_{SC-SFM}|$. The arrows show the 'time' evolution of the field profiles, and units for $\textbf{B}$ are $(\mu_{0}/4\pi)J_{c}R_{SC}$.
}
\end{center}
\end{figure*}

All the above can be seen in better detail from Fig.~\ref{Fig_3}, where we have displayed the local profiles for the norm of the magnetic field along two different radial directions, i.e., along the $(r,\phi)$ lines with $\phi=0$ and $\phi=\pi/4$, respectively, either (i) inside the SC wire $(0<r<1)$, (ii) inside the SFM $(1<r<1.5)$, or (iii) outside the SC-SFM metastructure $(1.5<r)$. The magnetic behavior of the SC-SFM  (solid lines) is compared with the classical critical-state behavior computed for an unsheathed SC wire (dashed lines), where besides the rapid rise of the magnetic field at the interface between the SC and the SFM, no disturbance of the magnetic field has been observed along the $y-$axis $(\phi=\pi/2)$, in good agreement with the experimental measurements for Fe $(\mu_{r}=46)$ sheathed MgB$_{2}$ wires~\cite{Pan2003,Pan2004,Roussel2006,Roussel2007}, where it has systematically reported an unusual ``elevation'' and ``dip'' of the magnetic flux only around the interface between the SC and the SFM when $\phi=0~(or ~\pi)$. In fact, the non-divergence and continuity conditions of the magnetic field can be directly observed in this figure, as the ``elevation'' in the flux free regions develops symmetrically from the condition $\phi=\pm\pi/2$ towards $\phi=0$, showing already a rise in the magnetic field along the line $(r,\phi/4)$ inside the SC (Fig.~\ref{Fig_3}~(a)) with the ``dip'' being evident at $\phi=0$ (Fig.~\ref{Fig_3}~(b)), either from the first ramp of the applied transport current $(0<\omega t<\pi/2)$, emulating the DC behavior, or during the hysteretic period shown in Figs.~\ref{Fig_3}~(d)-(e) for the angles $\phi=0$ and $\phi=\pi/4$, respectively. 

Then, as it is shown in Fig.~\ref{Fig_3}~(c) and Fig.~\ref{Fig_3}~(e), besides the rapid change in the intensity of the magnetic field that occurs at the interfaces between the SC and the SFM at $r=1$, and the SFM and the EXT domain at $r=1.5$, which are both caused by the change in the relative magnetic permeability of the medium, there is almost a negligible change in the slope or pattern of the magnetic field profile outside the SC-SFM metastructure at self-field conditions. Also, it is worth mentioning than the curves displayed in Fig.~\ref{Fig_3} refer directly to the calculations made within the numerical minimization framework of Eq.~\ref{Eq_3}, i.e., with the profiles of current density directly calculated by our numerical method, and then used to calculate the magnetic field from our analytical derivations at Eqs.~\ref{Eq_16}-\ref{Eq_18}. Therefore, the exact position where the sudden rise or drop of the magnetic field near the interfaces abovementioned is shown, can be somehow overestimated as it depends on the size of the finite elements considered for defining the local profiles of current density $J_{i}$. Still, such features have been experimentally observed from magneto optical imaging measurements~\cite{Pan2004,Roussel2006}, reason why for the purpose of proving the general validity of the critical state theory, these results will be qualitatively compared with our numerical observations in the following chapter.

%%%%%%%%%%%%%%%%%%%%%%%%%%%%%%%%%%%%%%%%%%%%%%%%%%%%%%%%%%%%%%%%%%%%%%%%%%%%%%%
%% FIGURE 3
%%%%%%%%%%%%%%%%%%%%%%%%%%%%%%%%%%%%%%%%%%%%%%%%%%%%%%%%%%%%%%%%%%%%%%%%%%%%%%%
%%%%%%%%%%%%%%%%%%%
%%%%%%%%%%%%%%%%%%%

%%%%%%%%%%%%%%%%%%%%%%%%%%%%%%%%%%%%%%%%%%%%%%%%%%%%%%%%%%%%%%%%%%%%%%%%%%%%%%%
%% SECTION 4
%%%%%%%%%%%%%%%%%%%%%%%%%%%%%%%%%%%%%%%%%%%%%%%%%%%%%%%%%%%%%%%%%%%%%%%%%%%%%%%

\section{Experimental evidences and general map of AC-Losses for SC-SFM Heterostructures}\label{Sec.4}

Visualizing and understanding the magnetic response ``inside'' of a SC-SFM metastructure, under transport current conditions, is undoubtedly a remarkable challenge either from the theoretical, computational, or experimental points of view. This is not only because the classical formulation of the CST and the solution of Maxwell equations are commonly do not include the physical coupling between these materials (as explained in Sec. 2) but, also because the experimental measurement of the local magnetic field at cryogenic temperatures is generally restricted to built on-purpose equipment, it tailored within already sophisticated experimental techniques. In this regard, although it is not our aim to discuss the diverse magnetic imaging techniques that could be used for this purpose, nor to provide a depth analysis of these~\cite{Miai_2001,Budker_2013,Jooss2002}, it is worthy of mentioning that there are two different visualization methods for the local imaging of magnetic fields in superconductors that stand out. The first of these methods corresponds to the use of polarized neutrons allowing to reveal the three-dimensional distribution of magnetic fields in solid materials~\cite{Kardjilov2011,Kardjilov2018}. This technique provides the best spatial resolution of all the local magnetic imaging techniques, but up to date, there are no reported measurements on monocore superconducting wires under transport current conditions, nor on comparable SC-SFM metastructures. Nevertheless, there is a second method of interest that corresponds to the so-called MOI technique~\cite{Roussel2007,Jooss2002}, from which the main observations reported for SC and SC-SFM cylindrical metastructures have been reproduced via our extended CST.

%%%%%%%%%%%%%%%%%%%%%%%%%%%%%%%%%%%%%%%%%%%%%%%%%%%%%%%%%%%%%%%%%%%%%%%%%%%%%%%
%% FIGURE 4
%%%%%%%%%%%%%%%%%%%%%%%%%%%%%%%%%%%%%%%%%%%%%%%%%%%%%%%%%%%%%%%%%%%%%%%%%%%%%%%
%%%%%%%%%%%%%%%%%%%
\begin{figure}[t]
\begin{center}
{\includegraphics[width=0.48\textwidth]{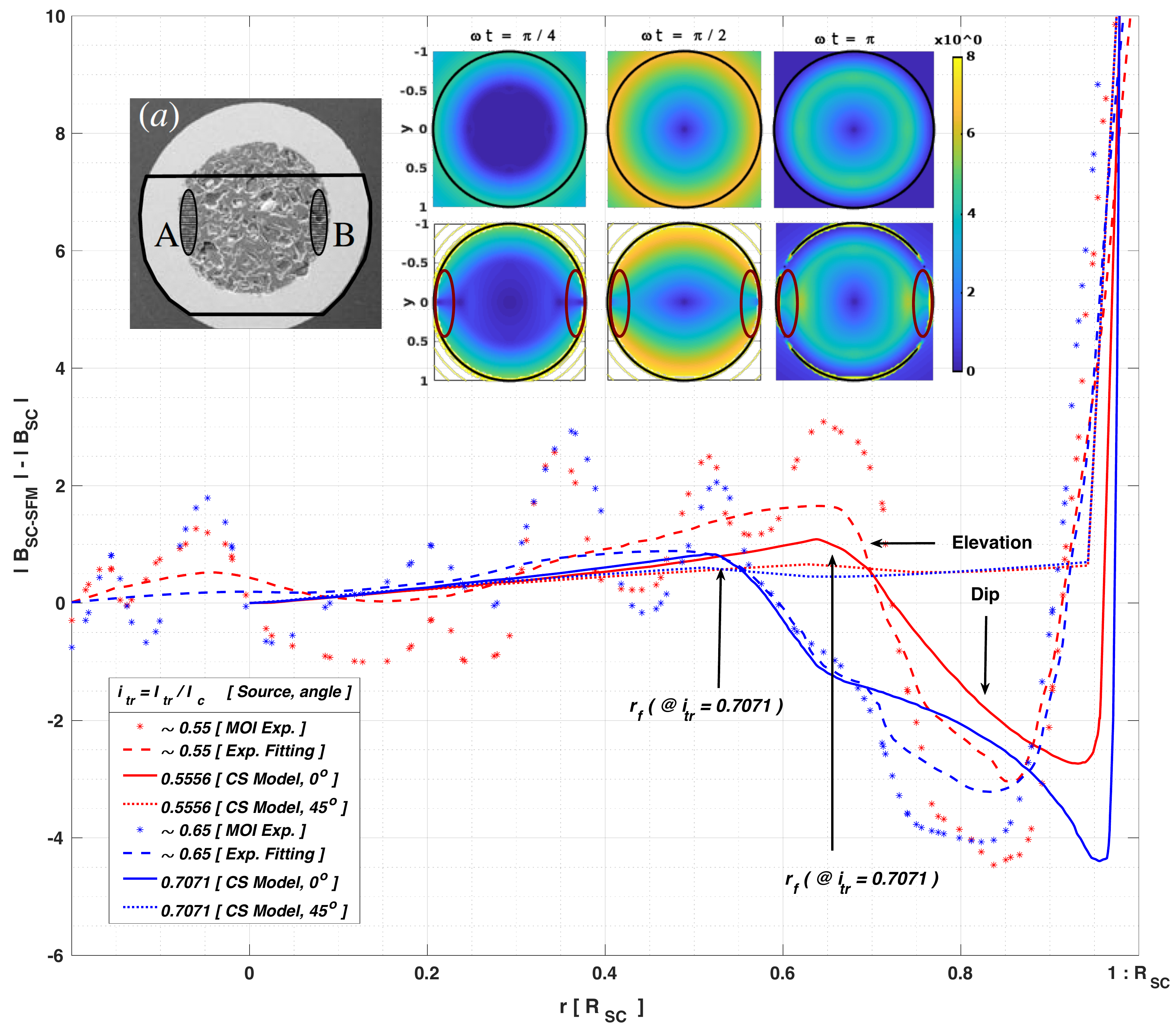}}
\caption{\label{Fig_4} Magnetic flux difference between a SC wire of radius $R_{SC}$ and the equivalent SC-SFM metastructure with $R_{SFM}=1.5R_{SC}$ in units of $(\mu_{0}/4\pi)J_{c}R_{SC}$. Solid and dotted lines show the numerical results obtained by the extended CST along two different radial directions, being $0^{\circ}$ the line over the x-axis at $y=0$ in Fig.~\ref{Fig_1}, and $45^{\circ}$ the $xy-$plane diagonal. Two set of curves are shown corresponding to self-field conditions with $I_{tr}=I_{c}\sin(\omega t)$, when $\omega t=3\pi/16$ and $\pi/4$ (i.e., $I_{tr}/I_{c}\simeq 0.5556$ and $0.7071$, respectively). For qualitative comparison, solid symbols and dashed lines show the raw and segmented-regression fitted data extracted from MOI measurements~\cite{Pan2003,Pan2004,Roussel2006,Roussel2007} reported for the Fe-sheathed \MgB2 monocore displayed at the subplot (a), where the ovals highlight the regions were an anomalous ``elevation'' and ``dip'' of the magnetic flux have been observed. All other insets show the calculated 2D local distribution of magnetic flux density at different instants of the AC current for the SC (top) and SC-SFM (bottom) wires. Equally sized ovals as in (a) are displayed, highlighting thence how the extended CST allows a straightforward explanation of the non-conventional patterns for the local magnetic flux density in the SC-SFM metastructures.
}
\end{center}
\end{figure}
%%%%%%%%%%%%%%%%%%%

In simple terms, the MOI technique makes use of the Faraday effect (sometimes called Faraday rotation) on birefringent doped ferrite garnet films with in-plane magnetization. This facilitates the real-time visualization of 2D magnetic field distributions of samples placed in close contact to the film, by detecting the rotation of the angle of a linearly polarized light beam above the magneto-optical film, which allows to determine the local distribution of magnetic field below it. By the use of this method, experimental measurements on the local distribution of magnetic field inside Fe-sheathed and bare \MgB2 wires have been previously reported~\cite{Pan2004,Pan2004a,Roussel2006,Roussel2007}, revealing some of the intriguing features for SC-SFM metastructures that have motivated this manuscript. These are, the occurrence of magnetic field within a region of the SC core that was expected to be ``flux-free'', leading to an ``elevation'' of the magnetic field profile around the center of the SC core, it towards a striking ``dip'' in the magnetic flux seen near the interface between the SC and the SFM materials.

These somehow exotic phenomena were originally thought to be caused by some mechanism similar to the overcritical state model in thin SC strips by Genenko et. al.~\cite{Genenko2002,Genenko2000PRB,Genenko2002PRB}, as from the classical perspective of Bean's model for the CST, the occurrence of local magnetic fields in the SC state, automatically means the occurrence of superconducting current densities. Therefore, the fact of having magnetic field at a region were no current density was to be expected, plus having localized areas where the magnetic self-field was even lower than what was seen by a bare SC (see the oval-shadowed areas in Fig.~\ref{Fig_2}(a)), could be intuitively explained by the occurrence of overcritical currents at those regions where the magnetic field has decreased. That, if we accept the counter-intuitive idea of having regions where violating the condition $|J| \leq |J_{c}|$ does not lead to the destruction of superconductivity.

Nevertheless, direct experimental measurements of the critical current density have shown no increment on $J_{c0}$ between the SC and SC-SFM wires under self-field conditions~\cite{Roussel2006,Roussel2007,Wells2011,Horvat2008}. However, contrary to think that these phenomena could imply a violation of the CST, we have demonstrated that the inclusion of the magnetic multipoles created by the SC-SFM coupling (see sec.~\ref{Sec.2}), are actually sufficient to reproduce all the macroscopic electromagnetic features seen by the MOI experiments. In this sense, in Fig.~\ref{Fig_2} we show the main electromagnetic features captured by the MOI technique, these against the theoretical predictions encompassed by our extended CST. There, it is to be noticed that despite having a clear proximity between the experimental and the theoretical results, a comparison between these data must be understood by, preferably, a qualitative rather than a quantitative manner. This is simply because of the limitations on the resolution encountered by the MOI measurements, which do not allow to have a straightforward quantitative comparison with our numerical predictions.

Thus, it is worth reminding that for a proper reading of the experimental results, it must be done by bearing in mind that the magnetic field profile obtained with the MOI technique (in arbitrary units), is indirectly measured by the relationship between the spontaneous magnetization vector of the ferrite film, also called the Magneto Optical Layer (MOL), and the rotation angle of the polarized light. This light is detected by a crossed polarizer and analyser of the light path, that is placed before and after crossing the MOL.  Then, besides the different optical components that can induce unwanted depolarizing effects when the light beam is reflected from or transmitted through them, the experimental measurements can be affected by a possible lack of homogeneity on the in-plane magnetization of the MOL, and also by any other possible defect on the contact between this and the measurement sample~\cite{Roussel2007}. In consequence, the computation of the magnetic field profile is not made over a sole cut line, says over the x-axis in the sample shown in Fig.~\ref{Fig_4}, but as a relative third-degree polynomial reconstruction over a 2D line of approximately $10~\mu$m wide. Thus, the intensity of the light is calibrated by subtracting a calibration image from the image to be quantified~\cite{Roussel2007}, i.e., in our case, subtracting the image of the measured SC wire before being placed within the SFM sheath, $|B_{SC}|$, from the image for the SC-SFM metastructure, $|B_{SC-SFM}|$. Then, the MOI calibration program return a precision ranging from 2 to 10 mT, which for an \MgB2 wire as shown in Fig.~\ref{Fig_4}, with an approximate radius $R_{SC} \approx 280~\mu$m, and $I_{c} \approx 13$~A, implies a minimum relative tolerance of approximately 1.35 field-units, they defined as $(\mu_{0}/4\pi)J_{c}R_{SC} \simeq 1.48$~mT, i.e., accounting for the minimum precision of $2$~mT ($\sim 1.35*1.48$~mT). Therefore, it is not strange to observe flux jumps within the MOI measurements as shown in Fig.~\ref{Fig_4} (solid symbols) but what is actually strange, is to recurrently see a certain ``elevation'' and ``dip'' on the magnetic field around the flux-free front boundary $(r_{f})$ and the interface between the SC and the SFM sheath~\cite{Pan2003,Pan2004,Roussel2006,Roussel2007}.

On the one hand, the above can be said in a different manner, by remembering that for currents below $I_{c}$, a flux-free core region below $r_{f}$ (see Eq.~\ref{Eq_5}) is expected to be ``seen'' under the simplified CST, either for the bare SC or the SFM sheathed SC. However, in the latter a clear ``elevation'' of the magnetic flux appears when contrasted against the magnetic signal of the bare SC, it contrary to the classical predictions of the CST. Thus, if the magneto-steady coupling between the SC and the SFM is not directly included, just as it is shown in the case of the bare SC at the top inset of Fig.~\ref{Fig_4}, when $\omega t = \pi/4$ (i.e., when $I_{tr} \simeq 0.7 I_{c}$), no difference between $|B_{SC}|$ and $|B_{SC-SFM}|$ should be seen for $r\lesssim 0.5 R_{SC}$, as no magnetic field is to be seen at this region. However, as it can be observed by the fitting curves for the MOI measurements (Fig.~\ref{Fig_4}), in the case of the SC-SFM metastructure a clear rise in the magnetic flux for $r<r_{f}$ appears (dashed lines), which can be explained by the extended CST reported in this paper (solid lines). Moreover, by the extended CST, we have proven that this anomalous rise in the magnetic flux within the ``flux-free core'' of the SC, a term brought up only for bare superconductors, is actually a direct consequence of the coupling between the SC and the SFM materials, i.e, it is the result of the magnetic multipoles induced by the interaction between the superconducting currents and the SFM sheath. In other words, in the case of a rounded SC-SFM metastructure at self field conditions, with either a DC or AC transport current of magnitude $I_{tr} < I_{c}$, the radius $r_{f}$ in Eq.~\ref{Eq_5} is simply referring to the boundary of a core free of transport current, but not necessarily free of magnetic flux. 

On the other hand, beyond the $r_{f}$ boundary, which in the case of the SC-SFM metastructure should be better called the transport current boundary, rather than the flux-front boundary, a pronounced ``dip'' on the magnetic field has been experimentally observed near the interface between the SC and the SFM materials. This unique magnetic feature of the SC-SFM metastructures which appears in a localized manner, i.e, only towards the sides of the SC wire as shown at the highlighted ovals in the insets of Fig.~\ref{Fig_4}, is indeed an even more anomalous feature in the distribution of magnetic field inside a SC, especially if it is seen from the point of view of the conventional CST. However, by means of our extended CST, these apparently anomalous features have been all duly reproduced in Fig.~\ref{Fig_4}, where the ``dip'' in the magnetic field profile can be clearly seen at $r>r_{f}$ for a cut-line at a $0^{\circ}$ polar angle (solid lines). Then, the magnetic flux intensity diminishes as the angle of measurement approaches to $\pm45^{\circ}$ (dotted lines), with a negligible difference being observed at $\pm90^{\circ}$ (not shown for the easy visualization of the other curves).

Likewise, it is worth mentioning that the rapid change in the intensity of the magnetic field that occurs at the interface between the SC and the SFM at $r=1$, and also, at the interface between the SFM and the EXT domain at $r=1.5$, both can be seen from the experimental or numerical points of view, either at Fig.~\ref{Fig_4}, or in greater detail at the Figures.~\ref{Fig_2} and \ref{Fig_3}. Thus, notice that this rise is just caused by the change in the relative magnetic permeability of the medium, and consequently by the continuity condition of the magnetic field, which applies regardless whether the CST has been extended or not. Then, outside of the SC-SFM metastructure, it has been found an almost negligible change in the slope or pattern of the magnetic field profile if compared with a bare SC, reaching nearly the same value of magnetic field at a distance less than just twice the radius of the SC-SFM wire.

%%%%%%%%%%%%%%%%%%%%%%%%%%%%%%%%%%%%%%%%%%%%%%%%%%%%%%%%%%%%%%%%%%%%%%%%%%%%%%%
%% FIGURE 5
%%%%%%%%%%%%%%%%%%%%%%%%%%%%%%%%%%%%%%%%%%%%%%%%%%%%%%%%%%%%%%%%%%%%%%%%%%%%%%%
%%%%%%%%%%%%%%%%%%%
\begin{figure}[t]
\begin{center}
{\includegraphics[width=0.48\textwidth]{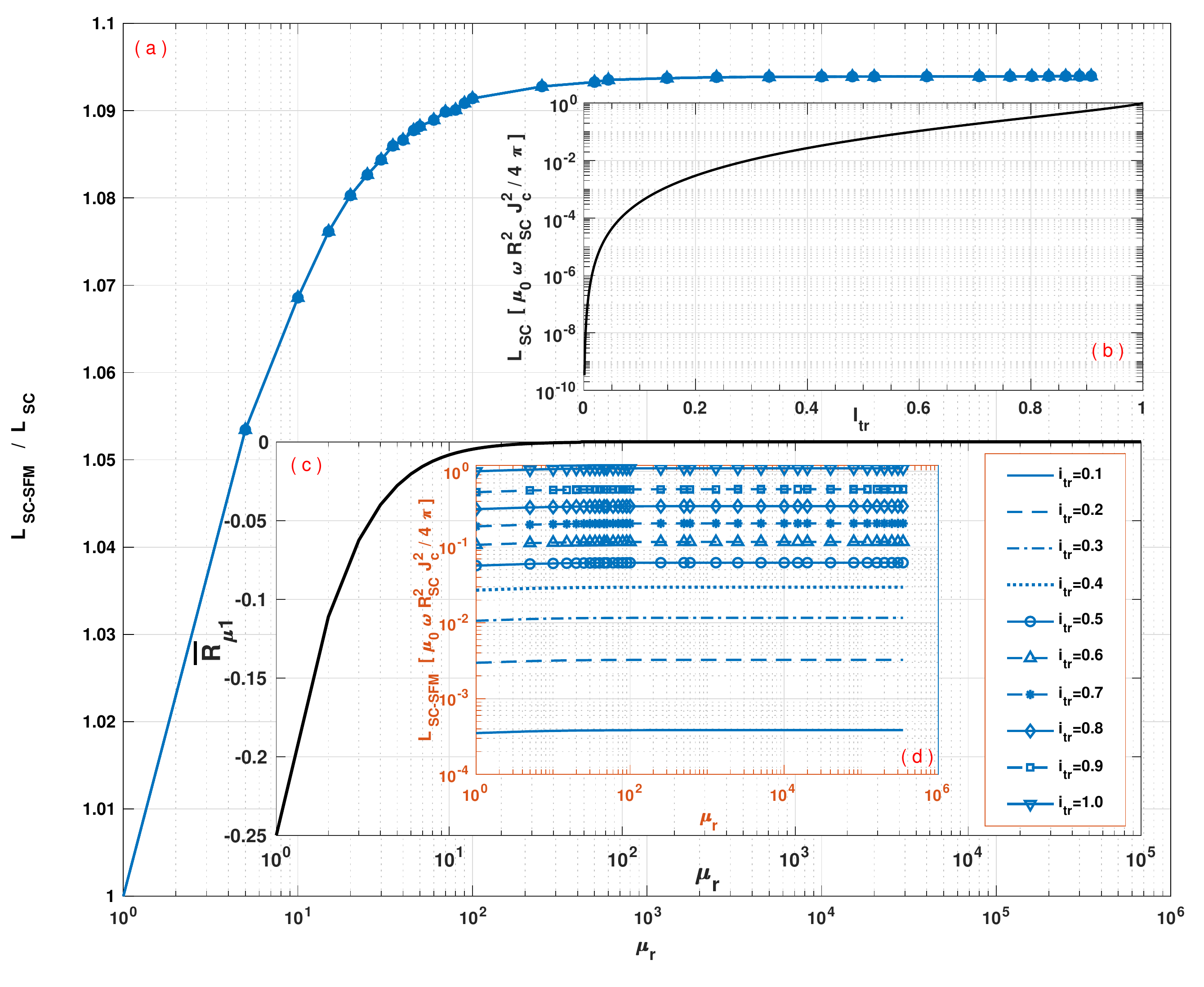}}
\caption{\label{Fig_5} (a) Hysteretic losses ratio between SC-SFM metastructures $(L_{SC-SFM})$ and the AC losses produced by an isolated SC wire $(L_{SC})$ of cylindrical cross section as a function of the relative magnetic permeability $\mu_{r}$ of the SFM with $R_{SFM}=1.5R_{SC}$. (b) The top inset shows the well known analytical solution for $L_{SC}$ as a function of $I_{tr}$~\cite{Gurevich1997,Ruiz2018bIEEE,Ruiz2012APL}. (c) The bottom outer inset shows the numerical tendency of the non-dimensional factor $\bar{R}_{\mu 1}$ (Eq.~\ref{Eq_16}) valid for any radius of the SFM sheath up to $R_{SFM}=10R_{SC}$ and with magnetic permeabilities from $\mu_{r}\sim1$ up to $1\times10^{5}$. (d) Finally, the inner inset shows the dependence of the metastructure losses  $(L_{SC-SFM})$ for different amplitudes of the transport current $i_{tr}$ in units of $I_{c}$, and the relative magnetic permeability of the SFM sheath $\mu_{r}$.
}
\end{center}
\end{figure}
%%%%%%%%%%%%%%%%%%%

Finally, the last relevant electromagnetic feature to be reproduced within the extended CST for SC-SFM wires under self-field conditions, concerns to what is probably the most important quantity to be measured within the framework of applied superconductivity, i.e. the AC-losses. In this sense, our extended CST also allows to prove that the coupling between the SC and the SFM sheath is indeed sufficient for explaining the slight rise seen in the AC-losses of SC-SFM wires~\cite{Eckelmann_1998,Huang_1998,Majoros2000}, even when the SFM layer does not add any electrical nor magnetic losses to the system (see Fig.~\ref{Fig_5}). Moreover, it has been found how the selection of the SFM magnetic properties, i.e., its relative magnetic permeability, can affect the most important observable macroscopic quantities such as the magnetic field created by the SC-SFM metastructure, and its energy losses.

To understand the impact of the SFM sheath on the hysteresis losses of SC-SFM metastructures, in Fig.~\ref{Fig_5} the curve of AC-losses for the rounded SC-SFM wire, $L_{SC-SFM}$, has been calculated as a function of the relative magnetic permeability of the SFM, $\mu_{r}$. The results are shown starting from the case of a bare SC, i.e., with $\mu_r=1$, up to an extremely high and rare magnetic permeability for SFM materials $(\mu_{r}=350000)$, which has been observed in pure Iron samples~\cite{Williams1992,DoD1987,Solymar1988}. This covers the whole range of SFM materials available in the market, such as Ni, NiZn, MnZn, Si, C, and Co ferrites $(\mu_{r}\cong 5-15000)$, providing the first known map of AC-losses for rounded SC-SFM wires. Thus, our study reveals a saturation of the hysteresis losses caused by the SFM at about $\mu_{r}=1000$, with even less than $1\%$ difference from $\mu_{r}=100$, but showing a very rapid change in the hysteresis curve between $\mu_{r}=1$ and $\mu_{r}=100$. This is a remarkable result, as it discloses that no matter the SFM used, beyond a relative magnetic permeability of $\mu_{r} \simeq 100$, nearly no increment of the AC-losses of the system could be observed by effect of the coupling between the SC and the SFM. 

Moreover, by analysing in detail the structure of the AC losses curve in Fig.~\ref{Fig_5}, it can be revealed that the dominant factor in the hysteresis losses provided by the coupling between the SC core and the SFM sheath, comes from the factor $\bar{R}_{\mu 1}$ in Eq.~\ref{Eq_13} (see Fig.~\ref{Fig_5}~(c)). Notice that this factor dominates also the contribution to the magnetic field outside of the SC-SFM metastructure in Eq.~\ref{Eq_20}, which appears as long as there is a locally induced profile of current $J_{i}$. Therefore, this might have strong implications in other phenomena such as the magnetic shielding and magnetic cloaking effects by SC-SFM metastructures, where an external magnetic field $\textbf{B}_{0}$ can be entered by the additional vector potential $\textbf{A}_{0}(\textbf{r}_{i})=\textbf{B}_{0}\times\textbf{r}_{i}$. The reason for this is because  these phenomena depend on the screening properties of the SC material, i.e, on the occurrence of local profiles of current density $\textbf{J}_{i}$, whose coupling with the SFM remains defined by the vector potentials for the couple media in Eqs.~\ref{Eq_13}-\ref{Eq_15}. Consequently, the shielding properties of the SFM in the rounded geometry of Fig.~\ref{Fig_1} are also limited by the factor $\bar{R}_{\mu 1}$, whose dependence on the relative radius between the SC and the SFM, as well as the magnetic permeability of the SFM, shown a nearly negligible impact for magnetic permeabilities greater than $\mu_{r}=100$, somehow contrary to the intuitive thinking that by increasing the magnetic permeability of the SFM sheath or its thickness on a SC cylindrical wire or tube, then the magnetic shielding properties of this heterostructure will increase too. 

%%%%%%%%%%%%%%%%%%%%%%%%%%%%%%%%%%%%%%%%%%%%%%%%%%%%%%%%%%%%%%%%%%%%%%%%%%%%%%%
%% SECTION 5
%%%%%%%%%%%%%%%%%%%%%%%%%%%%%%%%%%%%%%%%%%%%%%%%%%%%%%%%%%%%%%%%%%%%%%%%%%%%%%%

\section{Conclusions}~\label{Sec.5}

In this article, we have shown that the counterintuitive increment in the AC losses of monocore SC-SFM metastructures, at self-field conditions, can be explained under the conventional framework of the general critical state theory, without the need of recurring to the ansatz of overcritical currents. For doing so, the variational formulation of the CST introduced by Bad\'{\i}a, L\'{o}pez, and Ruiz~\cite{Ruiz2009bPRB}, has been extended in such way that the corresponding magnetic vector potentials are written within the magnetic multipole approach commonly used for accelerator magnets~\cite{Devred1998,Devred1999a,Devred1999b,Devred2005}.

In this way, we have proven that the intriguing anisotropy in the magnetic flux distribution inside the superconducting core of a SC-SFM heterostructure, results as a straightforward consequence of the magnetostatic coupling between these two materials. This coupling can be fully described by analytical methods within the conventional critical state framework for type-II superconductors, and without the need of considering the occurrence of edge currents, overcritical currents, or current sharing patterns which have not been experimentally observed at self-field conditions. Thus, as long as no current sharing between the SC and the SFM is enabled, i.e., when both materials are electrically insulated of each other, our semi-analytical model shows how despite there is a no alteration on the distribution of current density in the SC caused by the SFM sheath, yet it is possible to observe a certain amount of magnetic field in regions where no transport current is expected to flow. Likewise, we have shown how this magnetostatic coupling between the SC and the SFM materials, can cause the rather striking ``elevation'' and ``dip'' of the local magnetic field inside the core of SFM sheathed SC wires, which has been previously observed by Magneto Optical Imaging measurements~\cite{Pan2003,Pan2004a,Roussel2006}.

In addition, with the previous knowledge of the flux front profile for the current distribution inside a SC rod at self-field conditions, which cannot only be calculated by fully analytical methods, but also by the numerical minimization of our variational functional either with or without the SFM sheath, we have reported fully analytical solutions for the magnetic vector potential and the magnetic field vector at any region of the space for SC-SFM cylindrical heterostructures of arbitrary dimensions. With these, we provided a direct prove on the magneto-coupling physical mechanism that gives place to the field deformations inside SC-SFM heterostructures observed by MOI techniques~\cite{Pan2003,Pan2004,Roussel2006,Roussel2007}. Likewise, the cause of the intriguing increment in the AC-losses in SC-SFM heterostructures~\cite{Eckelmann_1998,Huang_1998,Majoros2000}, is explained as result of the found magneto-steady coupling between the SC and the SFM sheath. Thus, we have proven that the sole coupling between the SC and a SFM sheath is capable to produce the slight rise in the AC-losses at the SC material, even when the SFM does not add any electrical nor magnetic losses to the system. 

Finally, it has been found how the selection of the SFM magnetic properties, i.e., its relative magnetic permeability, can affect the most important observable macroscopic quantities such as the magnetic field created by the SC-SFM metastructure, and its energy losses. Remarkably, we have found that no matter the SFM used nor its dimensions, for relative magnetic permeabilities $\mu_{r}\gtrsim 100$, nearly no further increment on the AC-losses could be observed by effect of the coupling between the SC and the SFM, as the curve of losses is dominated by the  factor, $\bar{R}_{\mu 1}$ in Eq.~\ref{Eq_13}, disclosing in this way a comprehensive map of ac-losses for SC-SFM rounded heterostructures. Thence, the electromagnetic phenomena shown in this paper can be used as a benchmark to understand other technologies which can use SC-SFM rounded metastructures, such as in lossless three phase power cables and high energies accelerator magnets.

%%%%%%%%%%%%%%%%%%%%%%%%%%%%%%%%%%%%%%%%%%
\vspace{6pt} 
%%%%%%%%%%%%%%%%%%%%%%%%%%%%%%%%%%%%%%%%%%

%%%%%%%%%%%%%%%%%%%%%%%%%%%%%%%%%%%%%%%%%%

% Comment and/or uncomment the appropriate one at the point of submission

\section*{Funding}

This work was supported by the UK Research and Innovation, Engineering and Physical Sciences Research Council (EPSRC), through the grant Ref. EP/S025707/1 led by H.S.R

%%%%%%%%%%%%%%%%%%%%%%%%%%%%%%%%%%%%%%%%%%
\section*{Acknowledgments} 
All authors acknowledge the support of the East Midlands Energy Research Accelerator (ERA) and, the High Performance Computing Cluster Facility -ALICE- at the University of Leicester. MUF acknowledge the College of Science and Engineering Scholarship Unit of the University of Leicester. H.S.R. thanks A. Bad\'{\i}a-Majos, Y. Genenko and S. Yampolskii for the valuable discussions around SC-SFM metastructures.

\bibliography{References_Ruiz_Group}

\begin{thebibliography}{85}%
\makeatletter
\providecommand \@ifxundefined [1]{%
 \@ifx{#1\undefined}
}%
\providecommand \@ifnum [1]{%
 \ifnum #1\expandafter \@firstoftwo
 \else \expandafter \@secondoftwo
 \fi
}%
\providecommand \@ifx [1]{%
 \ifx #1\expandafter \@firstoftwo
 \else \expandafter \@secondoftwo
 \fi
}%
\providecommand \natexlab [1]{#1}%
\providecommand \enquote  [1]{``#1''}%
\providecommand \bibnamefont  [1]{#1}%
\providecommand \bibfnamefont [1]{#1}%
\providecommand \citenamefont [1]{#1}%
\providecommand \href@noop [0]{\@secondoftwo}%
\providecommand \href [0]{\begingroup \@sanitize@url \@href}%
\providecommand \@href[1]{\@@startlink{#1}\@@href}%
\providecommand \@@href[1]{\endgroup#1\@@endlink}%
\providecommand \@sanitize@url [0]{\catcode `\\12\catcode `\$12\catcode
  `\&12\catcode `\#12\catcode `\^12\catcode `\_12\catcode `\%12\relax}%
\providecommand \@@startlink[1]{}%
\providecommand \@@endlink[0]{}%
\providecommand \url  [0]{\begingroup\@sanitize@url \@url }%
\providecommand \@url [1]{\endgroup\@href {#1}{\urlprefix }}%
\providecommand \urlprefix  [0]{URL }%
\providecommand \Eprint [0]{\href }%
\providecommand \doibase [0]{http://dx.doi.org/}%
\providecommand \selectlanguage [0]{\@gobble}%
\providecommand \bibinfo  [0]{\@secondoftwo}%
\providecommand \bibfield  [0]{\@secondoftwo}%
\providecommand \translation [1]{[#1]}%
\providecommand \BibitemOpen [0]{}%
\providecommand \bibitemStop [0]{}%
\providecommand \bibitemNoStop [0]{.\EOS\space}%
\providecommand \EOS [0]{\spacefactor3000\relax}%
\providecommand \BibitemShut  [1]{\csname bibitem#1\endcsname}%
\let\auto@bib@innerbib\@empty
%</preamble>
\bibitem [{\citenamefont {Glowacki}\ and\ \citenamefont
  {Majoros}(2009)}]{Glowacki2009}%
  \BibitemOpen
  \bibfield  {author} {\bibinfo {author} {\bibfnamefont {B.}~\bibnamefont
  {Glowacki}}\ and\ \bibinfo {author} {\bibfnamefont {M.}~\bibnamefont
  {Majoros}},\ }\href {\doibase https://doi.org/10.1088/0953-8984/21/25/254206}
  {\bibfield  {journal} {\bibinfo  {journal} {Journal of Physics: Condensed
  Matter}\ }\textbf {\bibinfo {volume} {21}},\ \bibinfo {pages} {254206}
  (\bibinfo {year} {2009})}\BibitemShut {NoStop}%
\bibitem [{\citenamefont {Sanchez}\ \emph {et~al.}(2011)\citenamefont
  {Sanchez}, \citenamefont {Navau}, \citenamefont {Prat-Camps},\ and\
  \citenamefont {Chen}}]{Sanchez2011}%
  \BibitemOpen
  \bibfield  {author} {\bibinfo {author} {\bibfnamefont {A.}~\bibnamefont
  {Sanchez}}, \bibinfo {author} {\bibfnamefont {C.}~\bibnamefont {Navau}},
  \bibinfo {author} {\bibfnamefont {J.}~\bibnamefont {Prat-Camps}}, \ and\
  \bibinfo {author} {\bibfnamefont {D.-X.}\ \bibnamefont {Chen}},\ }\href
  {\doibase https://doi.org/10.1088/1367-2630/13/9/093034} {\bibfield
  {journal} {\bibinfo  {journal} {New Journal of Physics}\ }\textbf {\bibinfo
  {volume} {13}},\ \bibinfo {pages} {093034} (\bibinfo {year}
  {2011})}\BibitemShut {NoStop}%
\bibitem [{\citenamefont {Navau}\ \emph {et~al.}(2014)\citenamefont {Navau},
  \citenamefont {Prat-Camps}, \citenamefont {Romero-Isart}, \citenamefont
  {Cirac},\ and\ \citenamefont {Sanchez}}]{Navau2014}%
  \BibitemOpen
  \bibfield  {author} {\bibinfo {author} {\bibfnamefont {C.}~\bibnamefont
  {Navau}}, \bibinfo {author} {\bibfnamefont {J.}~\bibnamefont {Prat-Camps}},
  \bibinfo {author} {\bibfnamefont {O.}~\bibnamefont {Romero-Isart}}, \bibinfo
  {author} {\bibfnamefont {J.~I.}\ \bibnamefont {Cirac}}, \ and\ \bibinfo
  {author} {\bibfnamefont {A.}~\bibnamefont {Sanchez}},\ }\href {\doibase
  https://doi.org/10.1103/PhysRevLett.112.253901} {\bibfield  {journal}
  {\bibinfo  {journal} {Physical Review Letters}\ }\textbf {\bibinfo {volume}
  {112}},\ \bibinfo {pages} {253901} (\bibinfo {year} {2014})}\BibitemShut
  {NoStop}%
\bibitem [{\citenamefont {Solovyov}\ \emph {et~al.}(2015)\citenamefont
  {Solovyov}, \citenamefont {\v{S}ouc},\ and\ \citenamefont
  {G\"{o}m\"{o}ry}}]{Solovyov2015}%
  \BibitemOpen
  \bibfield  {author} {\bibinfo {author} {\bibfnamefont {M.}~\bibnamefont
  {Solovyov}}, \bibinfo {author} {\bibfnamefont {J.}~\bibnamefont {\v{S}ouc}},
  \ and\ \bibinfo {author} {\bibfnamefont {F.}~\bibnamefont {G\"{o}m\"{o}ry}},\
  }\href {\doibase https://doi.org/10.1109/TASC.2014.2376176} {\bibfield
  {journal} {\bibinfo  {journal} {IEEE Transactions on Applied
  Superconductivity}\ }\textbf {\bibinfo {volume} {25}},\ \bibinfo {pages} {1}
  (\bibinfo {year} {2015})}\BibitemShut {NoStop}%
\bibitem [{\citenamefont {Genenko}\ \emph {et~al.}(2015)\citenamefont
  {Genenko}, \citenamefont {Rauh},\ and\ \citenamefont {Kurdi}}]{Genenko2015}%
  \BibitemOpen
  \bibfield  {author} {\bibinfo {author} {\bibfnamefont {Y.~A.}\ \bibnamefont
  {Genenko}}, \bibinfo {author} {\bibfnamefont {H.}~\bibnamefont {Rauh}}, \
  and\ \bibinfo {author} {\bibfnamefont {S.}~\bibnamefont {Kurdi}},\ }\href
  {\doibase https://doi.org/10.1063/1.4922982} {\bibfield  {journal} {\bibinfo
  {journal} {Journal of Applied Physics}\ }\textbf {\bibinfo {volume} {117}},\
  \bibinfo {pages} {243909} (\bibinfo {year} {2015})}\BibitemShut {NoStop}%
\bibitem [{\citenamefont {G{\"o}m{\"o}ry}\ \emph {et~al.}(2012)\citenamefont
  {G{\"o}m{\"o}ry}, \citenamefont {Solovyov}, \citenamefont {{\v{S}}ouc},
  \citenamefont {Navau}, \citenamefont {Prat-Camps},\ and\ \citenamefont
  {Sanchez}}]{Gomory2012}%
  \BibitemOpen
  \bibfield  {author} {\bibinfo {author} {\bibfnamefont {F.}~\bibnamefont
  {G{\"o}m{\"o}ry}}, \bibinfo {author} {\bibfnamefont {M.}~\bibnamefont
  {Solovyov}}, \bibinfo {author} {\bibfnamefont {J.}~\bibnamefont
  {{\v{S}}ouc}}, \bibinfo {author} {\bibfnamefont {C.}~\bibnamefont {Navau}},
  \bibinfo {author} {\bibfnamefont {J.}~\bibnamefont {Prat-Camps}}, \ and\
  \bibinfo {author} {\bibfnamefont {A.}~\bibnamefont {Sanchez}},\ }\href
  {\doibase https://doi.org/10.1126/science.1218316} {\bibfield  {journal}
  {\bibinfo  {journal} {Science}\ }\textbf {\bibinfo {volume} {335}},\ \bibinfo
  {pages} {1466} (\bibinfo {year} {2012})}\BibitemShut {NoStop}%
\bibitem [{\citenamefont {G\"{o}m\"{o}ry}\ \emph {et~al.}(2015)\citenamefont
  {G\"{o}m\"{o}ry}, \citenamefont {Solovyov},\ and\ \citenamefont
  {\v{S}ouc}}]{Gomory2015}%
  \BibitemOpen
  \bibfield  {author} {\bibinfo {author} {\bibfnamefont {F.}~\bibnamefont
  {G\"{o}m\"{o}ry}}, \bibinfo {author} {\bibfnamefont {M.}~\bibnamefont
  {Solovyov}}, \ and\ \bibinfo {author} {\bibfnamefont {J.}~\bibnamefont
  {\v{S}ouc}},\ }\href {\doibase https://doi.org/10.1088/0953-2048/28/4/044001}
  {\bibfield  {journal} {\bibinfo  {journal} {Superconductor Science and
  Technology}\ }\textbf {\bibinfo {volume} {28}},\ \bibinfo {pages} {044001}
  (\bibinfo {year} {2015})}\BibitemShut {NoStop}%
\bibitem [{\citenamefont {\v{S}ouc}\ \emph {et~al.}(2016)\citenamefont
  {\v{S}ouc}, \citenamefont {Solovyov},\ and\ \citenamefont
  {G\"{o}m\"{o}ry}}]{Souc2016}%
  \BibitemOpen
  \bibfield  {author} {\bibinfo {author} {\bibfnamefont {J.}~\bibnamefont
  {\v{S}ouc}}, \bibinfo {author} {\bibfnamefont {M.}~\bibnamefont {Solovyov}},
  \ and\ \bibinfo {author} {\bibfnamefont {F.}~\bibnamefont {G\"{o}m\"{o}ry}},\
  }\href {\doibase https://doi.org/10.1063/1.4959581} {\bibfield  {journal}
  {\bibinfo  {journal} {Applied Physics Letters}\ }\textbf {\bibinfo {volume}
  {109}},\ \bibinfo {pages} {033507} (\bibinfo {year} {2016})}\BibitemShut
  {NoStop}%
\bibitem [{\citenamefont {{Pe\~{n}a-Roche}}\ \emph {et~al.}(2016)\citenamefont
  {{Pe\~{n}a-Roche}}, \citenamefont {Genenko},\ and\ \citenamefont
  {Bad\'{\i}a-Maj\'os}}]{Badia2016}%
  \BibitemOpen
  \bibfield  {author} {\bibinfo {author} {\bibfnamefont {J.}~\bibnamefont
  {{Pe\~{n}a-Roche}}}, \bibinfo {author} {\bibfnamefont {Y.~A.}\ \bibnamefont
  {Genenko}}, \ and\ \bibinfo {author} {\bibfnamefont {A.}~\bibnamefont
  {Bad\'{\i}a-Maj\'os}},\ }\href {\doibase https://doi.org/10.1063/1.4961672}
  {\bibfield  {journal} {\bibinfo  {journal} {Applied Physics Letters}\
  }\textbf {\bibinfo {volume} {109}},\ \bibinfo {pages} {092601} (\bibinfo
  {year} {2016})}\BibitemShut {NoStop}%
\bibitem [{\citenamefont {{Fareed}}\ \emph {et~al.}(2019)\citenamefont
  {{Fareed}}, \citenamefont {{Robert}},\ and\ \citenamefont
  {{Ruiz}}}]{Ruiz2019IEEE}%
  \BibitemOpen
  \bibfield  {author} {\bibinfo {author} {\bibfnamefont {M.~U.}\ \bibnamefont
  {{Fareed}}}, \bibinfo {author} {\bibfnamefont {B.~C.}\ \bibnamefont
  {{Robert}}}, \ and\ \bibinfo {author} {\bibfnamefont {H.~S.}\ \bibnamefont
  {{Ruiz}}},\ }\href {\doibase https://doi.org/10.1109/TASC.2019.2893896}
  {\bibfield  {journal} {\bibinfo  {journal} {IEEE Transactions on Applied
  Superconductivity}\ }\textbf {\bibinfo {volume} {29}},\ \bibinfo {pages}
  {5900705} (\bibinfo {year} {2019})}\BibitemShut {NoStop}%
\bibitem [{\citenamefont {Baghdadi}\ \emph {et~al.}(2018)\citenamefont
  {Baghdadi}, \citenamefont {Ruiz},\ and\ \citenamefont
  {Coombs}}]{Ruiz2018SciRep}%
  \BibitemOpen
  \bibfield  {author} {\bibinfo {author} {\bibfnamefont {M.}~\bibnamefont
  {Baghdadi}}, \bibinfo {author} {\bibfnamefont {H.~S.}\ \bibnamefont {Ruiz}},
  \ and\ \bibinfo {author} {\bibfnamefont {T.~A.}\ \bibnamefont {Coombs}},\
  }\href {\doibase https://doi.org/10.1038/s41598-018-19681-8} {\bibfield
  {journal} {\bibinfo  {journal} {Scientific Reports}\ }\textbf {\bibinfo
  {volume} {8}},\ \bibinfo {pages} {1342} (\bibinfo {year} {2018})}\BibitemShut
  {NoStop}%
\bibitem [{\citenamefont {Genenko}\ \emph {et~al.}(2004)\citenamefont
  {Genenko}, \citenamefont {Yampolskii},\ and\ \citenamefont
  {Pan}}]{Genenko2004APL}%
  \BibitemOpen
  \bibfield  {author} {\bibinfo {author} {\bibfnamefont {Y.~A.}\ \bibnamefont
  {Genenko}}, \bibinfo {author} {\bibfnamefont {S.~V.}\ \bibnamefont
  {Yampolskii}}, \ and\ \bibinfo {author} {\bibfnamefont {A.~V.}\ \bibnamefont
  {Pan}},\ }\href {\doibase https://doi.org/10.1063/1.1741036} {\bibfield
  {journal} {\bibinfo  {journal} {Applied Physics Letters}\ }\textbf {\bibinfo
  {volume} {84}},\ \bibinfo {pages} {3921} (\bibinfo {year}
  {2004})}\BibitemShut {NoStop}%
\bibitem [{\citenamefont {Pang}\ \emph {et~al.}(1981)\citenamefont {Pang},
  \citenamefont {Campbell},\ and\ \citenamefont {McLaren}}]{Pang1981}%
  \BibitemOpen
  \bibfield  {author} {\bibinfo {author} {\bibfnamefont {C.}~\bibnamefont
  {Pang}}, \bibinfo {author} {\bibfnamefont {A.}~\bibnamefont {Campbell}}, \
  and\ \bibinfo {author} {\bibfnamefont {P.}~\bibnamefont {McLaren}},\ }\href
  {\doibase https://doi.org/10.1109/TMAG.1981.1061025} {\bibfield  {journal}
  {\bibinfo  {journal} {IEEE Transactions on Magnetics}\ }\textbf {\bibinfo
  {volume} {17}},\ \bibinfo {pages} {134} (\bibinfo {year} {1981})}\BibitemShut
  {NoStop}%
\bibitem [{\citenamefont {Kirchmayr}(1996)}]{Kirchmayr1996}%
  \BibitemOpen
  \bibfield  {author} {\bibinfo {author} {\bibfnamefont {H.~R.}\ \bibnamefont
  {Kirchmayr}},\ }\href {\doibase https://doi.org/10.1088/0022-3727/29/11/007}
  {\bibfield  {journal} {\bibinfo  {journal} {Journal of Physics D: Applied
  Physics}\ }\textbf {\bibinfo {volume} {29}},\ \bibinfo {pages} {2763}
  (\bibinfo {year} {1996})}\BibitemShut {NoStop}%
\bibitem [{\citenamefont {Itoh}\ \emph {et~al.}(1996)\citenamefont {Itoh},
  \citenamefont {Mori},\ and\ \citenamefont {Minemoto}}]{Itoh1996}%
  \BibitemOpen
  \bibfield  {author} {\bibinfo {author} {\bibfnamefont {M.}~\bibnamefont
  {Itoh}}, \bibinfo {author} {\bibfnamefont {K.}~\bibnamefont {Mori}}, \ and\
  \bibinfo {author} {\bibfnamefont {T.}~\bibnamefont {Minemoto}},\ }\href
  {\doibase https://doi.org/10.1109/20.511407} {\bibfield  {journal} {\bibinfo
  {journal} {IEEE Transactions on Magnetics}\ }\textbf {\bibinfo {volume}
  {32}},\ \bibinfo {pages} {2605} (\bibinfo {year} {1996})}\BibitemShut
  {NoStop}%
\bibitem [{\citenamefont {Lousberg}\ \emph {et~al.}(2010)\citenamefont
  {Lousberg}, \citenamefont {Fagnard}, \citenamefont {Ausloos}, \citenamefont
  {Vanderbemden},\ and\ \citenamefont {Vanderheyden}}]{Lousberg2010}%
  \BibitemOpen
  \bibfield  {author} {\bibinfo {author} {\bibfnamefont {G.~P.}\ \bibnamefont
  {Lousberg}}, \bibinfo {author} {\bibfnamefont {J.}~\bibnamefont {Fagnard}},
  \bibinfo {author} {\bibfnamefont {M.}~\bibnamefont {Ausloos}}, \bibinfo
  {author} {\bibfnamefont {P.}~\bibnamefont {Vanderbemden}}, \ and\ \bibinfo
  {author} {\bibfnamefont {B.}~\bibnamefont {Vanderheyden}},\ }\href {\doibase
  https://doi.org/10.1109/TASC.2009.2036855} {\bibfield  {journal} {\bibinfo
  {journal} {IEEE Transactions on Applied Superconductivity}\ }\textbf
  {\bibinfo {volume} {20}},\ \bibinfo {pages} {33} (\bibinfo {year}
  {2010})}\BibitemShut {NoStop}%
\bibitem [{\citenamefont {Lousberg}\ \emph {et~al.}(2011)\citenamefont
  {Lousberg}, \citenamefont {Fagnard}, \citenamefont {Chaud}, \citenamefont
  {Ausloos}, \citenamefont {Vanderbemden},\ and\ \citenamefont
  {Vanderheyden}}]{Lousberg2011}%
  \BibitemOpen
  \bibfield  {author} {\bibinfo {author} {\bibfnamefont {G.~P.}\ \bibnamefont
  {Lousberg}}, \bibinfo {author} {\bibfnamefont {J.-F.}\ \bibnamefont
  {Fagnard}}, \bibinfo {author} {\bibfnamefont {X.}~\bibnamefont {Chaud}},
  \bibinfo {author} {\bibfnamefont {M.}~\bibnamefont {Ausloos}}, \bibinfo
  {author} {\bibfnamefont {P.}~\bibnamefont {Vanderbemden}}, \ and\ \bibinfo
  {author} {\bibfnamefont {B.}~\bibnamefont {Vanderheyden}},\ }\href {\doibase
  https://doi.org/10.1088/0953-2048/24/3/035008} {\bibfield  {journal}
  {\bibinfo  {journal} {Superconductor Science and Technology}\ }\textbf
  {\bibinfo {volume} {24}},\ \bibinfo {pages} {035008} (\bibinfo {year}
  {2011})}\BibitemShut {NoStop}%
\bibitem [{\citenamefont {Horvat}\ \emph {et~al.}(2008)\citenamefont {Horvat},
  \citenamefont {Yeoh}, \citenamefont {Kim},\ and\ \citenamefont
  {Dou}}]{Horvat2008}%
  \BibitemOpen
  \bibfield  {author} {\bibinfo {author} {\bibfnamefont {J.}~\bibnamefont
  {Horvat}}, \bibinfo {author} {\bibfnamefont {W.~K.}\ \bibnamefont {Yeoh}},
  \bibinfo {author} {\bibfnamefont {J.~H.}\ \bibnamefont {Kim}}, \ and\
  \bibinfo {author} {\bibfnamefont {S.~X.}\ \bibnamefont {Dou}},\ }\href
  {\doibase https://doi.org/10.1088/0953-2048/21/6/065003} {\bibfield
  {journal} {\bibinfo  {journal} {Superconductor Science and Technology}\
  }\textbf {\bibinfo {volume} {21}},\ \bibinfo {pages} {065003} (\bibinfo
  {year} {2008})}\BibitemShut {NoStop}%
\bibitem [{\citenamefont {Horvat}\ \emph {et~al.}(2005)\citenamefont {Horvat},
  \citenamefont {Soltanian},\ and\ \citenamefont {Yeoh}}]{Horvat_2005_SUST}%
  \BibitemOpen
  \bibfield  {author} {\bibinfo {author} {\bibfnamefont {J.}~\bibnamefont
  {Horvat}}, \bibinfo {author} {\bibfnamefont {S.}~\bibnamefont {Soltanian}}, \
  and\ \bibinfo {author} {\bibfnamefont {W.~K.}\ \bibnamefont {Yeoh}},\ }\href
  {\doibase https://doi.org/10.1088/0953-2048/18/5/017} {\bibfield  {journal}
  {\bibinfo  {journal} {Superconductor Science and Technology}\ }\textbf
  {\bibinfo {volume} {18}},\ \bibinfo {pages} {682} (\bibinfo {year}
  {2005})}\BibitemShut {NoStop}%
\bibitem [{\citenamefont {Kov\'{a}\v{c}}\ \emph {et~al.}(2006)\citenamefont
  {Kov\'{a}\v{c}}, \citenamefont {Hu\v{s}ek}, \citenamefont {Meli\v{s}ek},
  \citenamefont {Kulich},\ and\ \citenamefont {\v{S}trb\'{i}k}}]{Kovac2006}%
  \BibitemOpen
  \bibfield  {author} {\bibinfo {author} {\bibfnamefont {P.}~\bibnamefont
  {Kov\'{a}\v{c}}}, \bibinfo {author} {\bibfnamefont {I.}~\bibnamefont
  {Hu\v{s}ek}}, \bibinfo {author} {\bibfnamefont {T.}~\bibnamefont
  {Meli\v{s}ek}}, \bibinfo {author} {\bibfnamefont {M.}~\bibnamefont {Kulich}},
  \ and\ \bibinfo {author} {\bibfnamefont {V.}~\bibnamefont {\v{S}trb\'{i}k}},\
  }\href {\doibase https://doi.org/10.1088/0953-2048/19/6/031} {\bibfield
  {journal} {\bibinfo  {journal} {Superconductor Science and Technology}\
  }\textbf {\bibinfo {volume} {19}},\ \bibinfo {pages} {600} (\bibinfo {year}
  {2006})}\BibitemShut {NoStop}%
\bibitem [{\citenamefont {Gozzelino}\ \emph {et~al.}(2013)\citenamefont
  {Gozzelino}, \citenamefont {Gerbaldo}, \citenamefont {Ghigo}, \citenamefont
  {Laviano}, \citenamefont {Agostino}, \citenamefont {Bonometti}, \citenamefont
  {Chiampi}, \citenamefont {Manzin},\ and\ \citenamefont
  {Zilberti}}]{Gozzelino2013}%
  \BibitemOpen
  \bibfield  {author} {\bibinfo {author} {\bibfnamefont {L.}~\bibnamefont
  {Gozzelino}}, \bibinfo {author} {\bibfnamefont {R.}~\bibnamefont {Gerbaldo}},
  \bibinfo {author} {\bibfnamefont {G.}~\bibnamefont {Ghigo}}, \bibinfo
  {author} {\bibfnamefont {F.}~\bibnamefont {Laviano}}, \bibinfo {author}
  {\bibfnamefont {A.}~\bibnamefont {Agostino}}, \bibinfo {author}
  {\bibfnamefont {E.}~\bibnamefont {Bonometti}}, \bibinfo {author}
  {\bibfnamefont {M.}~\bibnamefont {Chiampi}}, \bibinfo {author} {\bibfnamefont
  {A.}~\bibnamefont {Manzin}}, \ and\ \bibinfo {author} {\bibfnamefont
  {L.}~\bibnamefont {Zilberti}},\ }\href {\doibase
  https://doi.org/10.1109/TASC.2012.2234817} {\bibfield  {journal} {\bibinfo
  {journal} {IEEE Transactions on Applied Superconductivity}\ }\textbf
  {\bibinfo {volume} {23}},\ \bibinfo {pages} {8201305} (\bibinfo {year}
  {2013})}\BibitemShut {NoStop}%
\bibitem [{\citenamefont {Gozzelino}\ \emph {et~al.}(2017)\citenamefont
  {Gozzelino}, \citenamefont {Gerbaldo}, \citenamefont {Ghigo}, \citenamefont
  {Laviano},\ and\ \citenamefont {Truccato}}]{Gozzelino2017}%
  \BibitemOpen
  \bibfield  {author} {\bibinfo {author} {\bibfnamefont {L.}~\bibnamefont
  {Gozzelino}}, \bibinfo {author} {\bibfnamefont {R.}~\bibnamefont {Gerbaldo}},
  \bibinfo {author} {\bibfnamefont {G.}~\bibnamefont {Ghigo}}, \bibinfo
  {author} {\bibfnamefont {F.}~\bibnamefont {Laviano}}, \ and\ \bibinfo
  {author} {\bibfnamefont {M.}~\bibnamefont {Truccato}},\ }\href {\doibase
  https://doi.org/10.1007/s10948-016-3659-z} {\bibfield  {journal} {\bibinfo
  {journal} {Journal of Superconductivity and Novel Magnetism}\ }\textbf
  {\bibinfo {volume} {30}},\ \bibinfo {pages} {749} (\bibinfo {year}
  {2017})}\BibitemShut {NoStop}%
\bibitem [{\citenamefont {Genenko}\ \emph
  {et~al.}(2000{\natexlab{a}})\citenamefont {Genenko}, \citenamefont
  {Snezhko},\ and\ \citenamefont {Freyhardt}}]{Genenko2000PRB}%
  \BibitemOpen
  \bibfield  {author} {\bibinfo {author} {\bibfnamefont {Y.~A.}\ \bibnamefont
  {Genenko}}, \bibinfo {author} {\bibfnamefont {A.}~\bibnamefont {Snezhko}}, \
  and\ \bibinfo {author} {\bibfnamefont {H.~C.}\ \bibnamefont {Freyhardt}},\
  }\href {\doibase https://doi.org/10.1103/PhysRevB.62.3453} {\bibfield
  {journal} {\bibinfo  {journal} {Phys. Rev. B}\ }\textbf {\bibinfo {volume}
  {62}},\ \bibinfo {pages} {3453} (\bibinfo {year}
  {2000}{\natexlab{a}})}\BibitemShut {NoStop}%
\bibitem [{\citenamefont {Genenko}\ \emph
  {et~al.}(2000{\natexlab{b}})\citenamefont {Genenko}, \citenamefont {Usoskin},
  \citenamefont {Snezhko},\ and\ \citenamefont {Freyhardt}}]{Genenko2000}%
  \BibitemOpen
  \bibfield  {author} {\bibinfo {author} {\bibfnamefont {Y.}~\bibnamefont
  {Genenko}}, \bibinfo {author} {\bibfnamefont {A.}~\bibnamefont {Usoskin}},
  \bibinfo {author} {\bibfnamefont {A.}~\bibnamefont {Snezhko}}, \ and\
  \bibinfo {author} {\bibfnamefont {H.}~\bibnamefont {Freyhardt}},\ }\href
  {\doibase https://doi.org/10.1016/S0921-4534(00)00784-X} {\bibfield
  {journal} {\bibinfo  {journal} {Physica C: Superconductivity}\ }\textbf
  {\bibinfo {volume} {341-348}},\ \bibinfo {pages} {1063 } (\bibinfo {year}
  {2000}{\natexlab{b}})}\BibitemShut {NoStop}%
\bibitem [{\citenamefont {Genenko}\ and\ \citenamefont
  {Snezhko}(2002)}]{Genenko2002}%
  \BibitemOpen
  \bibfield  {author} {\bibinfo {author} {\bibfnamefont {Y.~A.}\ \bibnamefont
  {Genenko}}\ and\ \bibinfo {author} {\bibfnamefont {A.~V.}\ \bibnamefont
  {Snezhko}},\ }\href {\doibase https://doi.org/10.1063/1.1480112} {\bibfield
  {journal} {\bibinfo  {journal} {Journal of Applied Physics}\ }\textbf
  {\bibinfo {volume} {92}},\ \bibinfo {pages} {357} (\bibinfo {year}
  {2002})}\BibitemShut {NoStop}%
\bibitem [{\citenamefont {Genenko}(2002{\natexlab{a}})}]{Genenko2002PRB}%
  \BibitemOpen
  \bibfield  {author} {\bibinfo {author} {\bibfnamefont {Y.~A.}\ \bibnamefont
  {Genenko}},\ }\href {\doibase https://doi.org/10.1103/PhysRevB.66.184520}
  {\bibfield  {journal} {\bibinfo  {journal} {Phys. Rev. B}\ }\textbf {\bibinfo
  {volume} {66}},\ \bibinfo {pages} {184520} (\bibinfo {year}
  {2002}{\natexlab{a}})}\BibitemShut {NoStop}%
\bibitem [{\citenamefont {Genenko}\ and\ \citenamefont
  {Rauh}(2003)}]{Genenko2003}%
  \BibitemOpen
  \bibfield  {author} {\bibinfo {author} {\bibfnamefont {Y.~A.}\ \bibnamefont
  {Genenko}}\ and\ \bibinfo {author} {\bibfnamefont {H.}~\bibnamefont {Rauh}},\
  }\href {\doibase https://doi.org/10.1063/1.1560866} {\bibfield  {journal}
  {\bibinfo  {journal} {Applied Physics Letters}\ }\textbf {\bibinfo {volume}
  {82}},\ \bibinfo {pages} {2115} (\bibinfo {year} {2003})}\BibitemShut
  {NoStop}%
\bibitem [{\citenamefont {Genenko}\ and\ \citenamefont
  {Rauh}(2006)}]{Genenko2006}%
  \BibitemOpen
  \bibfield  {author} {\bibinfo {author} {\bibfnamefont {Y.~A.}\ \bibnamefont
  {Genenko}}\ and\ \bibinfo {author} {\bibfnamefont {H.}~\bibnamefont {Rauh}},\
  }\href {\doibase https://doi.org/10.1088/1742-6596/43/1/140} {\bibfield
  {journal} {\bibinfo  {journal} {Journal of Physics: Conference Series}\
  }\textbf {\bibinfo {volume} {43}},\ \bibinfo {pages} {568} (\bibinfo {year}
  {2006})}\BibitemShut {NoStop}%
\bibitem [{\citenamefont {Yampolskii}\ \emph {et~al.}(2007)\citenamefont
  {Yampolskii}, \citenamefont {Genenko},\ and\ \citenamefont
  {Rauh}}]{Genenko2007}%
  \BibitemOpen
  \bibfield  {author} {\bibinfo {author} {\bibfnamefont {S.}~\bibnamefont
  {Yampolskii}}, \bibinfo {author} {\bibfnamefont {Y.}~\bibnamefont {Genenko}},
  \ and\ \bibinfo {author} {\bibfnamefont {H.}~\bibnamefont {Rauh}},\ }\href
  {\doibase https://doi.org/10.1016/j.physc.2007.04.065} {\bibfield  {journal}
  {\bibinfo  {journal} {Physica C: Superconductivity}\ }\textbf {\bibinfo
  {volume} {460-462}},\ \bibinfo {pages} {1262 } (\bibinfo {year} {2007})},\
  \bibinfo {note} {proceedings of the 8th International Conference on Materials
  and Mechanisms of Superconductivity and High Temperature
  Superconductors}\BibitemShut {NoStop}%
\bibitem [{\citenamefont {Genenko}\ \emph {et~al.}(2011)\citenamefont
  {Genenko}, \citenamefont {Rauh},\ and\ \citenamefont
  {Kr\"{u}ger}}]{Genenko2011}%
  \BibitemOpen
  \bibfield  {author} {\bibinfo {author} {\bibfnamefont {Y.~A.}\ \bibnamefont
  {Genenko}}, \bibinfo {author} {\bibfnamefont {H.}~\bibnamefont {Rauh}}, \
  and\ \bibinfo {author} {\bibfnamefont {P.}~\bibnamefont {Kr\"{u}ger}},\
  }\href {\doibase https://doi.org/10.1063/1.3560461} {\bibfield  {journal}
  {\bibinfo  {journal} {Applied Physics Letters}\ }\textbf {\bibinfo {volume}
  {98}},\ \bibinfo {pages} {152508} (\bibinfo {year} {2011})}\BibitemShut
  {NoStop}%
\bibitem [{\citenamefont {Genenko}(2002{\natexlab{b}})}]{Genenko2012}%
  \BibitemOpen
  \bibfield  {author} {\bibinfo {author} {\bibfnamefont {Y.~A.}\ \bibnamefont
  {Genenko}},\ }\href@noop {} {\bibfield  {journal} {\bibinfo  {journal}
  {Physica Status Solidi (a)}\ }\textbf {\bibinfo {volume} {189}},\ \bibinfo
  {pages} {469} (\bibinfo {year} {2002}{\natexlab{b}})}\BibitemShut {NoStop}%
\bibitem [{\citenamefont {Yampolskii}\ \emph {et~al.}(2004)\citenamefont
  {Yampolskii}, \citenamefont {Genenko},\ and\ \citenamefont
  {Rauh}}]{Genenko2004}%
  \BibitemOpen
  \bibfield  {author} {\bibinfo {author} {\bibfnamefont {S.}~\bibnamefont
  {Yampolskii}}, \bibinfo {author} {\bibfnamefont {Y.}~\bibnamefont {Genenko}},
  \ and\ \bibinfo {author} {\bibfnamefont {H.}~\bibnamefont {Rauh}},\ }\href
  {\doibase https://doi.org/10.1016/j.physc.2004.05.012} {\bibfield  {journal}
  {\bibinfo  {journal} {Physica C: Superconductivity}\ }\textbf {\bibinfo
  {volume} {415}},\ \bibinfo {pages} {151 } (\bibinfo {year}
  {2004})}\BibitemShut {NoStop}%
\bibitem [{\citenamefont {Genenko}\ \emph {et~al.}(2005)\citenamefont
  {Genenko}, \citenamefont {Rauh},\ and\ \citenamefont
  {Yampolskii}}]{Genenko2005}%
  \BibitemOpen
  \bibfield  {author} {\bibinfo {author} {\bibfnamefont {Y.~A.}\ \bibnamefont
  {Genenko}}, \bibinfo {author} {\bibfnamefont {H.}~\bibnamefont {Rauh}}, \
  and\ \bibinfo {author} {\bibfnamefont {S.~V.}\ \bibnamefont {Yampolskii}},\
  }\href {\doibase https://doi.org/10.1088/0953-8984/17/10/L02} {\bibfield
  {journal} {\bibinfo  {journal} {Journal of Physics: Condensed Matter}\
  }\textbf {\bibinfo {volume} {17}},\ \bibinfo {pages} {L93} (\bibinfo {year}
  {2005})}\BibitemShut {NoStop}%
\bibitem [{\citenamefont {Yampolskii}\ and\ \citenamefont
  {Genenko}(2005)}]{Genenko2005PRB}%
  \BibitemOpen
  \bibfield  {author} {\bibinfo {author} {\bibfnamefont {S.~V.}\ \bibnamefont
  {Yampolskii}}\ and\ \bibinfo {author} {\bibfnamefont {Y.~A.}\ \bibnamefont
  {Genenko}},\ }\href {\doibase https://doi.org/10.1103/PhysRevB.71.134519}
  {\bibfield  {journal} {\bibinfo  {journal} {Phys. Rev. B}\ }\textbf {\bibinfo
  {volume} {71}},\ \bibinfo {pages} {134519} (\bibinfo {year}
  {2005})}\BibitemShut {NoStop}%
\bibitem [{\citenamefont {Yampolskii}\ \emph {et~al.}(2006)\citenamefont
  {Yampolskii}, \citenamefont {Genenko}, \citenamefont {Rauh},\ and\
  \citenamefont {Snezhko}}]{Genenko2006JoP}%
  \BibitemOpen
  \bibfield  {author} {\bibinfo {author} {\bibfnamefont {S.~V.}\ \bibnamefont
  {Yampolskii}}, \bibinfo {author} {\bibfnamefont {Y.~A.}\ \bibnamefont
  {Genenko}}, \bibinfo {author} {\bibfnamefont {H.}~\bibnamefont {Rauh}}, \
  and\ \bibinfo {author} {\bibfnamefont {A.~V.}\ \bibnamefont {Snezhko}},\
  }\href {\doibase https://doi.org/10.1088/1742-6596/43/1/142} {\bibfield
  {journal} {\bibinfo  {journal} {Journal of Physics: Conference Series}\
  }\textbf {\bibinfo {volume} {43}},\ \bibinfo {pages} {576} (\bibinfo {year}
  {2006})}\BibitemShut {NoStop}%
\bibitem [{\citenamefont {Eckelmann}\ \emph {et~al.}(1998)\citenamefont
  {Eckelmann}, \citenamefont {D\''{a}umling}, \citenamefont {Quilitz},\ and\
  \citenamefont {Goldacker}}]{Eckelmann_1998}%
  \BibitemOpen
  \bibfield  {author} {\bibinfo {author} {\bibfnamefont {H.}~\bibnamefont
  {Eckelmann}}, \bibinfo {author} {\bibfnamefont {M.}~\bibnamefont
  {D\''{a}umling}}, \bibinfo {author} {\bibfnamefont {M.}~\bibnamefont
  {Quilitz}}, \ and\ \bibinfo {author} {\bibfnamefont {W.}~\bibnamefont
  {Goldacker}},\ }\href {\doibase
  https://doi.org/10.1016/S0921-4534(97)01773-5} {\bibfield  {journal}
  {\bibinfo  {journal} {Physica. C, Superconductivity}\ }\textbf {\bibinfo
  {volume} {295}},\ \bibinfo {pages} {198} (\bibinfo {year}
  {1998})}\BibitemShut {NoStop}%
\bibitem [{\citenamefont {Huang}\ \emph {et~al.}(1998)\citenamefont {Huang},
  \citenamefont {Dhall\'{e}}, \citenamefont {Witz}, \citenamefont {Marti},
  \citenamefont {Giannini}, \citenamefont {Walker}, \citenamefont {Passerini},
  \citenamefont {Polcari}, \citenamefont {Clerc}, \citenamefont {Kwasnitza},\
  and\ \citenamefont {Fl\''{u}kiger}}]{Huang_1998}%
  \BibitemOpen
  \bibfield  {author} {\bibinfo {author} {\bibfnamefont {Y.~B.}\ \bibnamefont
  {Huang}}, \bibinfo {author} {\bibfnamefont {M.}~\bibnamefont {Dhall\'{e}}},
  \bibinfo {author} {\bibfnamefont {G.}~\bibnamefont {Witz}}, \bibinfo {author}
  {\bibfnamefont {F.}~\bibnamefont {Marti}}, \bibinfo {author} {\bibfnamefont
  {E.}~\bibnamefont {Giannini}}, \bibinfo {author} {\bibfnamefont
  {E.}~\bibnamefont {Walker}}, \bibinfo {author} {\bibfnamefont
  {R.}~\bibnamefont {Passerini}}, \bibinfo {author} {\bibfnamefont
  {A.}~\bibnamefont {Polcari}}, \bibinfo {author} {\bibfnamefont
  {S.}~\bibnamefont {Clerc}}, \bibinfo {author} {\bibfnamefont
  {K.}~\bibnamefont {Kwasnitza}}, \ and\ \bibinfo {author} {\bibfnamefont
  {R.}~\bibnamefont {Fl\''{u}kiger}},\ }\href {\doibase
  https://doi.org/10.1023/A:1022662607835} {\bibfield  {journal} {\bibinfo
  {journal} {Journal of Superconductivity: Incorporating Novel Magnetism}\
  }\textbf {\bibinfo {volume} {11}},\ \bibinfo {pages} {495} (\bibinfo {year}
  {1998})}\BibitemShut {NoStop}%
\bibitem [{\citenamefont {Majoros}\ \emph {et~al.}(2000)\citenamefont
  {Majoros}, \citenamefont {Glowacki},\ and\ \citenamefont
  {Campbell}}]{Majoros2000}%
  \BibitemOpen
  \bibfield  {author} {\bibinfo {author} {\bibfnamefont {M.}~\bibnamefont
  {Majoros}}, \bibinfo {author} {\bibfnamefont {B.}~\bibnamefont {Glowacki}}, \
  and\ \bibinfo {author} {\bibfnamefont {A.}~\bibnamefont {Campbell}},\ }\href
  {\doibase https://doi.org/10.1016/S0921-4534(00)00276-8} {\bibfield
  {journal} {\bibinfo  {journal} {Physica C: Superconductivity}\ }\textbf
  {\bibinfo {volume} {334}},\ \bibinfo {pages} {129 } (\bibinfo {year}
  {2000})}\BibitemShut {NoStop}%
\bibitem [{\citenamefont {Pan}\ \emph {et~al.}(2003)\citenamefont {Pan},
  \citenamefont {Zhou}, \citenamefont {Liu},\ and\ \citenamefont
  {Dou}}]{Pan2003}%
  \BibitemOpen
  \bibfield  {author} {\bibinfo {author} {\bibfnamefont {A.~V.}\ \bibnamefont
  {Pan}}, \bibinfo {author} {\bibfnamefont {S.}~\bibnamefont {Zhou}}, \bibinfo
  {author} {\bibfnamefont {H.}~\bibnamefont {Liu}}, \ and\ \bibinfo {author}
  {\bibfnamefont {S.}~\bibnamefont {Dou}},\ }\href {\doibase
  https://doi.org/10.1088/0953-2048/16/10/101} {\bibfield  {journal} {\bibinfo
  {journal} {Superconductor Science and Technology}\ }\textbf {\bibinfo
  {volume} {16}},\ \bibinfo {pages} {L33} (\bibinfo {year} {2003})}\BibitemShut
  {NoStop}%
\bibitem [{\citenamefont {Pan}\ \emph {et~al.}(2004)\citenamefont {Pan},
  \citenamefont {Dou},\ and\ \citenamefont {H.}}]{Pan2004a}%
  \BibitemOpen
  \bibfield  {author} {\bibinfo {author} {\bibfnamefont {A.~V.}\ \bibnamefont
  {Pan}}, \bibinfo {author} {\bibfnamefont {S.}~\bibnamefont {Dou}}, \ and\
  \bibinfo {author} {\bibfnamefont {J.~T.}\ \bibnamefont {H.}},\ }in\ \href
  {\doibase https://doi.org/10.1007/978-94-007-1007-8_18} {\emph {\bibinfo
  {booktitle} {Magneto-Optical Imaging. NATO Science Series (Series II:
  Mathematics, Physics and Chemistry)}}},\ Vol.\ \bibinfo {volume} {142}\
  (\bibinfo  {publisher} {Springer},\ \bibinfo {address} {Dordrecht},\ \bibinfo
  {year} {2004})\BibitemShut {NoStop}%
\bibitem [{\citenamefont {Roussel}\ \emph {et~al.}(2006)\citenamefont
  {Roussel}, \citenamefont {Pan}, \citenamefont {Zhou},\ and\ \citenamefont
  {Dou}}]{Roussel2006}%
  \BibitemOpen
  \bibfield  {author} {\bibinfo {author} {\bibfnamefont {M.}~\bibnamefont
  {Roussel}}, \bibinfo {author} {\bibfnamefont {A.~V.}\ \bibnamefont {Pan}},
  \bibinfo {author} {\bibfnamefont {S.}~\bibnamefont {Zhou}}, \ and\ \bibinfo
  {author} {\bibfnamefont {S.~X.}\ \bibnamefont {Dou}},\ }\href {\doibase
  https://doi.org/10.1088/1742-6596/43/1/024} {\bibfield  {journal} {\bibinfo
  {journal} {Journal of Physics: Conference Series}\ }\textbf {\bibinfo
  {volume} {43}},\ \bibinfo {pages} {95} (\bibinfo {year} {2006})}\BibitemShut
  {NoStop}%
\bibitem [{\citenamefont {Horvat}\ \emph {et~al.}(2002)\citenamefont {Horvat},
  \citenamefont {Wang}, \citenamefont {Soltanian},\ and\ \citenamefont
  {Dou}}]{Horvat_2002_APL}%
  \BibitemOpen
  \bibfield  {author} {\bibinfo {author} {\bibfnamefont {J.}~\bibnamefont
  {Horvat}}, \bibinfo {author} {\bibfnamefont {X.~L.}\ \bibnamefont {Wang}},
  \bibinfo {author} {\bibfnamefont {S.}~\bibnamefont {Soltanian}}, \ and\
  \bibinfo {author} {\bibfnamefont {S.~X.}\ \bibnamefont {Dou}},\ }\href
  {\doibase https://doi.org/10.1063/1.1447010} {\bibfield  {journal} {\bibinfo
  {journal} {Applied Physics Letters}\ }\textbf {\bibinfo {volume} {80}},\
  \bibinfo {pages} {829} (\bibinfo {year} {2002})}\BibitemShut {NoStop}%
\bibitem [{\citenamefont {Roussel}(2007)}]{Roussel2007}%
  \BibitemOpen
  \bibfield  {author} {\bibinfo {author} {\bibfnamefont {M.}~\bibnamefont
  {Roussel}},\ }\href {http://ro.uow.edu.au/theses/16/} {\emph {\bibinfo
  {title} {Magneto-optical imaging in superconductors, PhD Thesis}}}\ (\bibinfo
   {publisher} {Faculty of Engineering, University of Wollongong Theses},\
  \bibinfo {address} {Wollongong, Australia},\ \bibinfo {year}
  {2007})\BibitemShut {NoStop}%
\bibitem [{\citenamefont {Wells}(2011)}]{Wells2011}%
  \BibitemOpen
  \bibfield  {author} {\bibinfo {author} {\bibfnamefont {F.~S.}\ \bibnamefont
  {Wells}},\ }\href {https://ro.uow.edu.au/theses/4351/} {\emph {\bibinfo
  {title} {Magneto-optical imaging and current profiling on superconductors,
  PhD Thesis}}}\ (\bibinfo  {publisher} {Faculty of Engineering, University of
  Wollongong Theses},\ \bibinfo {address} {Wollongong, Australia},\ \bibinfo
  {year} {2011})\BibitemShut {NoStop}%
\bibitem [{\citenamefont {Bad\'{\i}a-Maj\'os}\ \emph
  {et~al.}(2009)\citenamefont {Bad\'{\i}a-Maj\'os}, \citenamefont {L\'opez},\
  and\ \citenamefont {Ruiz}}]{Ruiz2009bPRB}%
  \BibitemOpen
  \bibfield  {author} {\bibinfo {author} {\bibfnamefont {A.}~\bibnamefont
  {Bad\'{\i}a-Maj\'os}}, \bibinfo {author} {\bibfnamefont {C.}~\bibnamefont
  {L\'opez}}, \ and\ \bibinfo {author} {\bibfnamefont {H.~S.}\ \bibnamefont
  {Ruiz}},\ }\href {\doibase https://doi.org/10.1103/PhysRevB.80.144509}
  {\bibfield  {journal} {\bibinfo  {journal} {Phys. Rev. B}\ }\textbf {\bibinfo
  {volume} {80}},\ \bibinfo {pages} {144509} (\bibinfo {year}
  {2009})}\BibitemShut {NoStop}%
\bibitem [{\citenamefont {Ruiz}\ \emph
  {et~al.}(2011{\natexlab{a}})\citenamefont {Ruiz}, \citenamefont
  {Bad\'{\i}a-Maj\'{o}s},\ and\ \citenamefont {L\'{o}pez}}]{Ruiz2011SUST}%
  \BibitemOpen
  \bibfield  {author} {\bibinfo {author} {\bibfnamefont {H.~S.}\ \bibnamefont
  {Ruiz}}, \bibinfo {author} {\bibfnamefont {A.}~\bibnamefont
  {Bad\'{\i}a-Maj\'{o}s}}, \ and\ \bibinfo {author} {\bibfnamefont
  {C.}~\bibnamefont {L\'{o}pez}},\ }\href {\doibase
  https://doi.org/10.1088/0953-2048/24/11/115005} {\bibfield  {journal}
  {\bibinfo  {journal} {Superconductor Science and Technology}\ }\textbf
  {\bibinfo {volume} {24}},\ \bibinfo {pages} {115005} (\bibinfo {year}
  {2011}{\natexlab{a}})}\BibitemShut {NoStop}%
\bibitem [{\citenamefont {Ruiz}\ \emph {et~al.}(2012)\citenamefont {Ruiz},
  \citenamefont {Bad\'{\i}a-Maj\'{o}s}, \citenamefont {Genenko}, \citenamefont
  {Rauh},\ and\ \citenamefont {Yampolskii}}]{Ruiz2012APL}%
  \BibitemOpen
  \bibfield  {author} {\bibinfo {author} {\bibfnamefont {H.~S.}\ \bibnamefont
  {Ruiz}}, \bibinfo {author} {\bibfnamefont {A.}~\bibnamefont
  {Bad\'{\i}a-Maj\'{o}s}}, \bibinfo {author} {\bibfnamefont {Y.~A.}\
  \bibnamefont {Genenko}}, \bibinfo {author} {\bibfnamefont {H.}~\bibnamefont
  {Rauh}}, \ and\ \bibinfo {author} {\bibfnamefont {S.~V.}\ \bibnamefont
  {Yampolskii}},\ }\href {\doibase https://doi.org/10.1063/1.3693614}
  {\bibfield  {journal} {\bibinfo  {journal} {Applied Physics Letters}\
  }\textbf {\bibinfo {volume} {100}},\ \bibinfo {pages} {112602} (\bibinfo
  {year} {2012})}\BibitemShut {NoStop}%
\bibitem [{\citenamefont {Ruiz}\ \emph {et~al.}(2013)\citenamefont {Ruiz},
  \citenamefont {Bad{\'{\i}}a-Maj{\'{o}}s}, \citenamefont {Genenko},\ and\
  \citenamefont {Yampolskii}}]{Ruiz2013IEEE}%
  \BibitemOpen
  \bibfield  {author} {\bibinfo {author} {\bibfnamefont {H.~S.}\ \bibnamefont
  {Ruiz}}, \bibinfo {author} {\bibfnamefont {A.}~\bibnamefont
  {Bad{\'{\i}}a-Maj{\'{o}}s}}, \bibinfo {author} {\bibfnamefont {Y.~A.}\
  \bibnamefont {Genenko}}, \ and\ \bibinfo {author} {\bibfnamefont {S.~V.}\
  \bibnamefont {Yampolskii}},\ }\href {\doibase
  https://doi.org/10.1109/TASC.2012.2232695} {\bibfield  {journal} {\bibinfo
  {journal} {IEEE Transactions on Applied Superconductivity}\ }\textbf
  {\bibinfo {volume} {23}},\ \bibinfo {pages} {8000404} (\bibinfo {year}
  {2013})}\BibitemShut {NoStop}%
\bibitem [{\citenamefont {Ruiz}\ and\ \citenamefont
  {Bad\'{\i}a-Maj\'{o}s}(2013)}]{Ruiz2013JAP}%
  \BibitemOpen
  \bibfield  {author} {\bibinfo {author} {\bibfnamefont {H.~S.}\ \bibnamefont
  {Ruiz}}\ and\ \bibinfo {author} {\bibfnamefont {A.}~\bibnamefont
  {Bad\'{\i}a-Maj\'{o}s}},\ }\href {\doibase https://doi.org/10.1063/1.4804931}
  {\bibfield  {journal} {\bibinfo  {journal} {Journal of Applied Physics}\
  }\textbf {\bibinfo {volume} {113}},\ \bibinfo {pages} {193906} (\bibinfo
  {year} {2013})}\BibitemShut {NoStop}%
\bibitem [{\citenamefont {Robert}\ and\ \citenamefont
  {Ruiz}(2018{\natexlab{a}})}]{Ruiz2018aIEEE}%
  \BibitemOpen
  \bibfield  {author} {\bibinfo {author} {\bibfnamefont {B.~C.}\ \bibnamefont
  {Robert}}\ and\ \bibinfo {author} {\bibfnamefont {H.~S.}\ \bibnamefont
  {Ruiz}},\ }\href {\doibase https://doi.org/10.1109/TASC.2017.2668060}
  {\bibfield  {journal} {\bibinfo  {journal} {IEEE Transactions on Applied
  Superconductivity}\ }\textbf {\bibinfo {volume} {28}},\ \bibinfo {pages}
  {8200905} (\bibinfo {year} {2018}{\natexlab{a}})}\BibitemShut {NoStop}%
\bibitem [{\citenamefont {Robert}\ and\ \citenamefont
  {Ruiz}(2018{\natexlab{b}})}]{Ruiz2018bIEEE}%
  \BibitemOpen
  \bibfield  {author} {\bibinfo {author} {\bibfnamefont {B.~C.}\ \bibnamefont
  {Robert}}\ and\ \bibinfo {author} {\bibfnamefont {H.~S.}\ \bibnamefont
  {Ruiz}},\ }\href {\doibase https://doi.org/10.1109/TASC.2018.2794138}
  {\bibfield  {journal} {\bibinfo  {journal} {IEEE Transactions on Applied
  Superconductivity}\ }\textbf {\bibinfo {volume} {28}},\ \bibinfo {pages}
  {8200805} (\bibinfo {year} {2018}{\natexlab{b}})}\BibitemShut {NoStop}%
\bibitem [{\citenamefont {Robert}\ and\ \citenamefont
  {Ruiz}(2018{\natexlab{c}})}]{Ruiz2018SUST}%
  \BibitemOpen
  \bibfield  {author} {\bibinfo {author} {\bibfnamefont {B.~C.}\ \bibnamefont
  {Robert}}\ and\ \bibinfo {author} {\bibfnamefont {H.~S.}\ \bibnamefont
  {Ruiz}},\ }\href {\doibase https://doi.org/10.1088/1361-6668/aaa823}
  {\bibfield  {journal} {\bibinfo  {journal} {Superconductor Science and
  Technology}\ } (\bibinfo {year} {2018}{\natexlab{c}}),\
  https://doi.org/10.1088/1361-6668/aaa823}\BibitemShut {NoStop}%
\bibitem [{\citenamefont {Robert}\ \emph {et~al.}(2019)\citenamefont {Robert},
  \citenamefont {Fareed},\ and\ \citenamefont {Ruiz}}]{Ruiz2019MDPI}%
  \BibitemOpen
  \bibfield  {author} {\bibinfo {author} {\bibfnamefont {B.~C.}\ \bibnamefont
  {Robert}}, \bibinfo {author} {\bibfnamefont {M.~U.}\ \bibnamefont {Fareed}},
  \ and\ \bibinfo {author} {\bibfnamefont {H.~S.}\ \bibnamefont {Ruiz}},\
  }\href {\doibase https://doi.org/10.3390/ma12172679} {\bibfield  {journal}
  {\bibinfo  {journal} {Materials}\ }\textbf {\bibinfo {volume} {12}},\
  \bibinfo {pages} {2679} (\bibinfo {year} {2019})}\BibitemShut {NoStop}%
\bibitem [{\citenamefont {Kov\'{a}\v{c}}\ \emph {et~al.}(2003)\citenamefont
  {Kov\'{a}\v{c}}, \citenamefont {Hu\v{s}ek}, \citenamefont {Meli\v{s}ek},
  \citenamefont {Ahoranta}, \citenamefont {\v{S}ouc}, \citenamefont
  {Lehtonen},\ and\ \citenamefont {G\"{o}m\"{o}ry}}]{Kovac_2003_SUST}%
  \BibitemOpen
  \bibfield  {author} {\bibinfo {author} {\bibfnamefont {P.}~\bibnamefont
  {Kov\'{a}\v{c}}}, \bibinfo {author} {\bibfnamefont {I.}~\bibnamefont
  {Hu\v{s}ek}}, \bibinfo {author} {\bibfnamefont {T.}~\bibnamefont
  {Meli\v{s}ek}}, \bibinfo {author} {\bibfnamefont {M.}~\bibnamefont
  {Ahoranta}}, \bibinfo {author} {\bibfnamefont {J.}~\bibnamefont {\v{S}ouc}},
  \bibinfo {author} {\bibfnamefont {J.}~\bibnamefont {Lehtonen}}, \ and\
  \bibinfo {author} {\bibfnamefont {F.}~\bibnamefont {G\"{o}m\"{o}ry}},\ }\href
  {\doibase https://doi.org/10.1088/0953-2048/16/10/312} {\bibfield  {journal}
  {\bibinfo  {journal} {Superconductor Science and Technology}\ }\textbf
  {\bibinfo {volume} {16}},\ \bibinfo {pages} {1195} (\bibinfo {year}
  {2003})}\BibitemShut {NoStop}%
\bibitem [{\citenamefont {Young}\ \emph {et~al.}(2007)\citenamefont {Young},
  \citenamefont {Bianchetti}, \citenamefont {Grasso},\ and\ \citenamefont
  {Yang}}]{Young_2007_IEEE}%
  \BibitemOpen
  \bibfield  {author} {\bibinfo {author} {\bibfnamefont {E.}~\bibnamefont
  {Young}}, \bibinfo {author} {\bibfnamefont {M.}~\bibnamefont {Bianchetti}},
  \bibinfo {author} {\bibfnamefont {G.}~\bibnamefont {Grasso}}, \ and\ \bibinfo
  {author} {\bibfnamefont {Y.}~\bibnamefont {Yang}},\ }\href {\doibase
  https://doi.org/10.1109/TASC.2007.899094} {\bibfield  {journal} {\bibinfo
  {journal} {IEEE Transactions on Applied Superconductivity}\ }\textbf
  {\bibinfo {volume} {17}},\ \bibinfo {pages} {2945} (\bibinfo {year}
  {2007})}\BibitemShut {NoStop}%
\bibitem [{\citenamefont {Majoros}\ \emph {et~al.}(2009)\citenamefont
  {Majoros}, \citenamefont {Sumption}, \citenamefont {Susner}, \citenamefont
  {Tomsic}, \citenamefont {Rindfleisch},\ and\ \citenamefont
  {Collings}}]{Majoros_2009_IEEE}%
  \BibitemOpen
  \bibfield  {author} {\bibinfo {author} {\bibfnamefont {M.}~\bibnamefont
  {Majoros}}, \bibinfo {author} {\bibfnamefont {M.}~\bibnamefont {Sumption}},
  \bibinfo {author} {\bibfnamefont {M.}~\bibnamefont {Susner}}, \bibinfo
  {author} {\bibfnamefont {M.}~\bibnamefont {Tomsic}}, \bibinfo {author}
  {\bibfnamefont {M.}~\bibnamefont {Rindfleisch}}, \ and\ \bibinfo {author}
  {\bibfnamefont {E.}~\bibnamefont {Collings}},\ }\href {\doibase
  https://doi.org/10.1109/TASC.2009.2019107} {\bibfield  {journal} {\bibinfo
  {journal} {IEEE Transactions on Applied Superconductivity}\ }\textbf
  {\bibinfo {volume} {19}},\ \bibinfo {pages} {3106} (\bibinfo {year}
  {2009})}\BibitemShut {NoStop}%
\bibitem [{\citenamefont {Nikulshin}\ \emph {et~al.}(2018)\citenamefont
  {Nikulshin}, \citenamefont {Wolfus}, \citenamefont {Friedman}, \citenamefont
  {Ginodman}, \citenamefont {Grasso}, \citenamefont {Tropeano}, \citenamefont
  {Bovone}, \citenamefont {Vignolo}, \citenamefont {Ferdeghini},\ and\
  \citenamefont {Yeshurun}}]{Nikulshin_2018_IEEE}%
  \BibitemOpen
  \bibfield  {author} {\bibinfo {author} {\bibfnamefont {Y.}~\bibnamefont
  {Nikulshin}}, \bibinfo {author} {\bibfnamefont {S.}~\bibnamefont {Wolfus}},
  \bibinfo {author} {\bibfnamefont {A.}~\bibnamefont {Friedman}}, \bibinfo
  {author} {\bibfnamefont {V.}~\bibnamefont {Ginodman}}, \bibinfo {author}
  {\bibfnamefont {G.}~\bibnamefont {Grasso}}, \bibinfo {author} {\bibfnamefont
  {M.}~\bibnamefont {Tropeano}}, \bibinfo {author} {\bibfnamefont
  {G.}~\bibnamefont {Bovone}}, \bibinfo {author} {\bibfnamefont
  {M.}~\bibnamefont {Vignolo}}, \bibinfo {author} {\bibfnamefont
  {C.}~\bibnamefont {Ferdeghini}}, \ and\ \bibinfo {author} {\bibfnamefont
  {Y.}~\bibnamefont {Yeshurun}},\ }\href {\doibase
  https://doi.org/10.1109/TASC.2018.2841926} {\bibfield  {journal} {\bibinfo
  {journal} {IEEE Transactions on Applied Superconductivity}\ }\textbf
  {\bibinfo {volume} {28}},\ \bibinfo {pages} {1} (\bibinfo {year}
  {2018})}\BibitemShut {NoStop}%
\bibitem [{\citenamefont {Nikulshin}\ \emph {et~al.}(2019)\citenamefont
  {Nikulshin}, \citenamefont {Yeshurun},\ and\ \citenamefont
  {Wolfus}}]{Nikulshin_2019_SUST}%
  \BibitemOpen
  \bibfield  {author} {\bibinfo {author} {\bibfnamefont {Y.}~\bibnamefont
  {Nikulshin}}, \bibinfo {author} {\bibfnamefont {Y.}~\bibnamefont {Yeshurun}},
  \ and\ \bibinfo {author} {\bibfnamefont {S.}~\bibnamefont {Wolfus}},\ }\href
  {\doibase https://doi.org/10.1088/1361-6668/ab13d9} {\bibfield  {journal}
  {\bibinfo  {journal} {Superconductor Science and Technology}\ }\textbf
  {\bibinfo {volume} {32}},\ \bibinfo {pages} {75007} (\bibinfo {year}
  {2019})}\BibitemShut {NoStop}%
\bibitem [{\citenamefont {Xi}\ \emph {et~al.}(2019)\citenamefont {Xi},
  \citenamefont {Pei}, \citenamefont {Sheng}, \citenamefont {Tanaka},
  \citenamefont {Ichiki}, \citenamefont {Zhang},\ and\ \citenamefont
  {Yuan}}]{Xi_2019_IEEE}%
  \BibitemOpen
  \bibfield  {author} {\bibinfo {author} {\bibfnamefont {J.}~\bibnamefont
  {Xi}}, \bibinfo {author} {\bibfnamefont {X.}~\bibnamefont {Pei}}, \bibinfo
  {author} {\bibfnamefont {J.}~\bibnamefont {Sheng}}, \bibinfo {author}
  {\bibfnamefont {H.}~\bibnamefont {Tanaka}}, \bibinfo {author} {\bibfnamefont
  {Y.}~\bibnamefont {Ichiki}}, \bibinfo {author} {\bibfnamefont
  {M.}~\bibnamefont {Zhang}}, \ and\ \bibinfo {author} {\bibfnamefont
  {W.}~\bibnamefont {Yuan}},\ }\href {\doibase
  https://doi.org/10.1109/TASC.2019.2903924} {\bibfield  {journal} {\bibinfo
  {journal} {IEEE Transactions on Applied Superconductivity}\ }\textbf
  {\bibinfo {volume} {29}},\ \bibinfo {pages} {1} (\bibinfo {year}
  {2019})}\BibitemShut {NoStop}%
\bibitem [{\citenamefont {Gurevich}\ \emph {et~al.}(1997)\citenamefont
  {Gurevich}, \citenamefont {Mints},\ and\ \citenamefont
  {Rakhmanov}}]{Gurevich1997}%
  \BibitemOpen
  \bibfield  {author} {\bibinfo {author} {\bibfnamefont {A.~V.}\ \bibnamefont
  {Gurevich}}, \bibinfo {author} {\bibfnamefont {R.~G.}\ \bibnamefont {Mints}},
  \ and\ \bibinfo {author} {\bibfnamefont {A.~L.}\ \bibnamefont {Rakhmanov}},\
  }\href@noop {} {\emph {\bibinfo {title} {Physics of Composite
  Superconductors}}}\ (\bibinfo  {publisher} {Begell House, New York},\
  \bibinfo {year} {1997})\ p.\ \bibinfo {pages} {348}\BibitemShut {NoStop}%
\bibitem [{\citenamefont {Fiorillo}\ and\ \citenamefont
  {Beatrice}(2011)}]{Fiorillo2011}%
  \BibitemOpen
  \bibfield  {author} {\bibinfo {author} {\bibfnamefont {F.}~\bibnamefont
  {Fiorillo}}\ and\ \bibinfo {author} {\bibfnamefont {C.}~\bibnamefont
  {Beatrice}},\ }\href {\doibase https://doi.org/10.1007/s10948-010-1000-9}
  {\bibfield  {journal} {\bibinfo  {journal} {Journal of Superconductivity and
  Novel Magnetism}\ }\textbf {\bibinfo {volume} {24}},\ \bibinfo {pages} {559}
  (\bibinfo {year} {2011})}\BibitemShut {NoStop}%
\bibitem [{\citenamefont {Coffey}\ and\ \citenamefont
  {Clem}(1991)}]{Coffey_1992_PRL}%
  \BibitemOpen
  \bibfield  {author} {\bibinfo {author} {\bibfnamefont {M.~W.}\ \bibnamefont
  {Coffey}}\ and\ \bibinfo {author} {\bibfnamefont {J.~R.}\ \bibnamefont
  {Clem}},\ }\href {\doibase 10.1103/PhysRevLett.67.386} {\bibfield  {journal}
  {\bibinfo  {journal} {Phys. Rev. Lett.}\ }\textbf {\bibinfo {volume} {67}},\
  \bibinfo {pages} {386} (\bibinfo {year} {1991})}\BibitemShut {NoStop}%
\bibitem [{\citenamefont {Coffey}(1993)}]{Coffey_1992}%
  \BibitemOpen
  \bibfield  {author} {\bibinfo {author} {\bibfnamefont {M.~W.}\ \bibnamefont
  {Coffey}},\ }\href {\doibase https://doi.org/10.1103/PhysRevB.47.12284}
  {\bibfield  {journal} {\bibinfo  {journal} {Phys. Rev. B}\ }\textbf {\bibinfo
  {volume} {47}},\ \bibinfo {pages} {12284} (\bibinfo {year}
  {1993})}\BibitemShut {NoStop}%
\bibitem [{\citenamefont {Coffey}\ and\ \citenamefont
  {Clem}(1992)}]{CoffeyClem_1992}%
  \BibitemOpen
  \bibfield  {author} {\bibinfo {author} {\bibfnamefont {M.~W.}\ \bibnamefont
  {Coffey}}\ and\ \bibinfo {author} {\bibfnamefont {J.~R.}\ \bibnamefont
  {Clem}},\ }\href {\doibase https://doi.org/10.1103/PhysRevB.46.11757}
  {\bibfield  {journal} {\bibinfo  {journal} {Phys. Rev. B}\ }\textbf {\bibinfo
  {volume} {46}},\ \bibinfo {pages} {11757} (\bibinfo {year}
  {1992})}\BibitemShut {NoStop}%
\bibitem [{\citenamefont {Grilli}\ \emph {et~al.}(2014)\citenamefont {Grilli},
  \citenamefont {Pardo}, \citenamefont {Stenvall}, \citenamefont {Nguyen},
  \citenamefont {Yuan},\ and\ \citenamefont {G\"{o}m\"ory}}]{Grilli2014}%
  \BibitemOpen
  \bibfield  {author} {\bibinfo {author} {\bibfnamefont {F.}~\bibnamefont
  {Grilli}}, \bibinfo {author} {\bibfnamefont {E.}~\bibnamefont {Pardo}},
  \bibinfo {author} {\bibfnamefont {A.}~\bibnamefont {Stenvall}}, \bibinfo
  {author} {\bibfnamefont {D.~N.}\ \bibnamefont {Nguyen}}, \bibinfo {author}
  {\bibfnamefont {W.}~\bibnamefont {Yuan}}, \ and\ \bibinfo {author}
  {\bibfnamefont {F.}~\bibnamefont {G\"{o}m\"ory}},\ }\href {\doibase
  https://doi.org/10.1109/TASC.2013.2259827} {\bibfield  {journal} {\bibinfo
  {journal} {IEEE Transactions on Applied Superconductivity}\ }\textbf
  {\bibinfo {volume} {24}},\ \bibinfo {pages} {78} (\bibinfo {year}
  {2014})}\BibitemShut {NoStop}%
\bibitem [{\citenamefont {Grilli}(2016)}]{Grilli2016}%
  \BibitemOpen
  \bibfield  {author} {\bibinfo {author} {\bibfnamefont {F.}~\bibnamefont
  {Grilli}},\ }\href {\doibase https://doi.org/10.1109/TASC.2016.2520083}
  {\bibfield  {journal} {\bibinfo  {journal} {IEEE Transactions on Applied
  Superconductivity}\ }\textbf {\bibinfo {volume} {26}},\ \bibinfo {pages}
  {0500408} (\bibinfo {year} {2016})}\BibitemShut {NoStop}%
\bibitem [{\citenamefont {Devred}(1998)}]{Devred1998}%
  \BibitemOpen
  \bibfield  {author} {\bibinfo {author} {\bibfnamefont {A.}~\bibnamefont
  {Devred}},\ }in\ \href@noop {} {\emph {\bibinfo {booktitle} {CERN Accelerator
  School (CAS 97): Measurement and Alignment of Accelerator and Detector
  Magnets}}}\ (\bibinfo {year} {1998})\ pp.\ \bibinfo {pages}
  {43--78}\BibitemShut {NoStop}%
\bibitem [{\citenamefont {Devred}(1999{\natexlab{a}})}]{Devred1999a}%
  \BibitemOpen
  \bibfield  {author} {\bibinfo {author} {\bibfnamefont {A.}~\bibnamefont
  {Devred}},\ }\href {\doibase https://doi.org/10.1002/047134608X.W1320} {\emph
  {\bibinfo {title} {Superconducting magnets for particle accelerators and
  storage rings}}}\ (\bibinfo  {publisher} {Wiley Encyclopedia of Electrical
  and Electronic Engineering},\ \bibinfo {year} {1999})\ p.\ \bibinfo {pages}
  {332}\BibitemShut {NoStop}%
\bibitem [{\citenamefont {Devred}(1999{\natexlab{b}})}]{Devred1999b}%
  \BibitemOpen
  \bibfield  {author} {\bibinfo {author} {\bibfnamefont {A.}~\bibnamefont
  {Devred}},\ }in\ \href@noop {} {\emph {\bibinfo {booktitle} {CEA Saclay,
  Dept. d'Astrophysique de la Physique des Particules de la Physique Nucleaire
  et de l'Instrumentation Associee (DAPNIA)}}}\ (\bibinfo {year} {1999})\ p.\
  \bibinfo {pages} {162}\BibitemShut {NoStop}%
\bibitem [{\citenamefont {Devred}\ and\ \citenamefont
  {Trassart}(2005)}]{Devred2005}%
  \BibitemOpen
  \bibfield  {author} {\bibinfo {author} {\bibfnamefont {A.}~\bibnamefont
  {Devred}}\ and\ \bibinfo {author} {\bibfnamefont {D.}~\bibnamefont
  {Trassart}},\ }in\ \href@noop {} {\emph {\bibinfo {booktitle} {European
  Organization for Nuclear Research, Laboratory for Particle Physics}}}\
  (\bibinfo {year} {2005})\BibitemShut {NoStop}%
\bibitem [{\citenamefont {Bad\'{\i}a}\ and\ \citenamefont
  {L\'opez}(2001)}]{Badia2001}%
  \BibitemOpen
  \bibfield  {author} {\bibinfo {author} {\bibfnamefont {A.}~\bibnamefont
  {Bad\'{\i}a}}\ and\ \bibinfo {author} {\bibfnamefont {C.}~\bibnamefont
  {L\'opez}},\ }\href {\doibase https://doi.org/10.1103/PhysRevLett.87.127004}
  {\bibfield  {journal} {\bibinfo  {journal} {Phys. Rev. Lett.}\ }\textbf
  {\bibinfo {volume} {87}},\ \bibinfo {pages} {127004} (\bibinfo {year}
  {2001})}\BibitemShut {NoStop}%
\bibitem [{\citenamefont {Bad\'{\i}a}\ and\ \citenamefont
  {L\'opez}(2002)}]{Badia2002}%
  \BibitemOpen
  \bibfield  {author} {\bibinfo {author} {\bibfnamefont {A.}~\bibnamefont
  {Bad\'{\i}a}}\ and\ \bibinfo {author} {\bibfnamefont {C.}~\bibnamefont
  {L\'opez}},\ }\href {\doibase https://doi.org/10.1103/PhysRevB.65.104514}
  {\bibfield  {journal} {\bibinfo  {journal} {Phys. Rev. B}\ }\textbf {\bibinfo
  {volume} {65}},\ \bibinfo {pages} {104514} (\bibinfo {year}
  {2002})}\BibitemShut {NoStop}%
\bibitem [{\citenamefont {Vlasko-Vlasov}\ \emph {et~al.}(2015)\citenamefont
  {Vlasko-Vlasov}, \citenamefont {Palacious}, \citenamefont {Rosenmann},
  \citenamefont {Pearson}, \citenamefont {Jia}, \citenamefont {Wang},
  \citenamefont {Welp},\ and\ \citenamefont {Kwok}}]{Vlasko2015}%
  \BibitemOpen
  \bibfield  {author} {\bibinfo {author} {\bibfnamefont {V.~K.}\ \bibnamefont
  {Vlasko-Vlasov}}, \bibinfo {author} {\bibfnamefont {E.}~\bibnamefont
  {Palacious}}, \bibinfo {author} {\bibfnamefont {D.}~\bibnamefont
  {Rosenmann}}, \bibinfo {author} {\bibfnamefont {J.}~\bibnamefont {Pearson}},
  \bibinfo {author} {\bibfnamefont {Y.}~\bibnamefont {Jia}}, \bibinfo {author}
  {\bibfnamefont {Y.~L.}\ \bibnamefont {Wang}}, \bibinfo {author}
  {\bibfnamefont {U.}~\bibnamefont {Welp}}, \ and\ \bibinfo {author}
  {\bibfnamefont {W.-K.}\ \bibnamefont {Kwok}},\ }\href {\doibase
  https://doi.org/10.1088/0953-2048/28/3/035006} {\bibfield  {journal}
  {\bibinfo  {journal} {Superconductor Science and Technology}\ }\textbf
  {\bibinfo {volume} {28}},\ \bibinfo {pages} {035006} (\bibinfo {year}
  {2015})}\BibitemShut {NoStop}%
\bibitem [{\citenamefont {Bean}(1962)}]{Bean1962}%
  \BibitemOpen
  \bibfield  {author} {\bibinfo {author} {\bibfnamefont {C.~P.}\ \bibnamefont
  {Bean}},\ }\href {\doibase https://doi.org/10.1103/PhysRevLett.8.250}
  {\bibfield  {journal} {\bibinfo  {journal} {Phys. Rev. Lett.}\ }\textbf
  {\bibinfo {volume} {8}},\ \bibinfo {pages} {250} (\bibinfo {year}
  {1962})}\BibitemShut {NoStop}%
\bibitem [{\citenamefont {Ruiz}\ \emph
  {et~al.}(2011{\natexlab{b}})\citenamefont {Ruiz}, \citenamefont {L\'opez},\
  and\ \citenamefont {Bad\'{\i}a-Maj\'os}}]{Ruiz2011PRB}%
  \BibitemOpen
  \bibfield  {author} {\bibinfo {author} {\bibfnamefont {H.~S.}\ \bibnamefont
  {Ruiz}}, \bibinfo {author} {\bibfnamefont {C.}~\bibnamefont {L\'opez}}, \
  and\ \bibinfo {author} {\bibfnamefont {A.}~\bibnamefont
  {Bad\'{\i}a-Maj\'os}},\ }\href {\doibase
  https://doi.org/10.1103/PhysRevB.83.014506} {\bibfield  {journal} {\bibinfo
  {journal} {Phys. Rev. B}\ }\textbf {\bibinfo {volume} {83}},\ \bibinfo
  {pages} {014506} (\bibinfo {year} {2011}{\natexlab{b}})}\BibitemShut
  {NoStop}%
\bibitem [{\citenamefont {Ruiz~Rondan}(2013)}]{Ruiz2013Book}%
  \BibitemOpen
  \bibfield  {author} {\bibinfo {author} {\bibfnamefont {H.}~\bibnamefont
  {Ruiz~Rondan}},\ }\href@noop {} {\emph {\bibinfo {title} {Material laws and
  numerical methods in applied superconductivity}}}\ (\bibinfo  {publisher}
  {University of Zaragoza Press},\ \bibinfo {address} {Zaragoza, Spain},\
  \bibinfo {year} {2013})\BibitemShut {NoStop}%
\bibitem [{\citenamefont {Williams}(1992)}]{Williams1992}%
  \BibitemOpen
  \bibfield  {author} {\bibinfo {author} {\bibfnamefont {B.}~\bibnamefont
  {Williams}},\ }\href {https://books.google.co.uk/books?id=I44eAQAAIAAJ}
  {\emph {\bibinfo {title} {Power Electronics: Devices, Drivers, Applications,
  and Passive Components}}}\ (\bibinfo  {publisher} {McGraw-Hill},\ \bibinfo
  {year} {1992})\BibitemShut {NoStop}%
\bibitem [{\citenamefont {DOD}(1987)}]{DoD1987}%
  \BibitemOpen
  \bibfield  {author} {\bibinfo {author} {\bibnamefont {DOD}},\ }\href
  {https://books.google.co.uk/books?id=I44eAQAAIAAJ} {\emph {\bibinfo {title}
  {Military Handbook. Grounding, Bonding, and Shielding for Electronic
  Equipments and Facilities}}},\ \bibinfo {type} {Tech. Rep.}\ (\bibinfo
  {institution} {Department of Defense, Washington, DC.},\ \bibinfo {year}
  {1987})\ \bibinfo {note} {mIL-HDBK-419A}\BibitemShut {NoStop}%
\bibitem [{\citenamefont {Solymar}\ \emph {et~al.}(1988)\citenamefont
  {Solymar}, \citenamefont {Solymar}, \citenamefont {Walsh},\ and\
  \citenamefont {Walsh}}]{Solymar1988}%
  \BibitemOpen
  \bibfield  {author} {\bibinfo {author} {\bibfnamefont {L.}~\bibnamefont
  {Solymar}}, \bibinfo {author} {\bibfnamefont {P.}~\bibnamefont {Solymar}},
  \bibinfo {author} {\bibfnamefont {D.}~\bibnamefont {Walsh}}, \ and\ \bibinfo
  {author} {\bibfnamefont {B.}~\bibnamefont {Walsh}},\ }\href
  {https://books.google.co.uk/books?id=oqHvAAAAMAAJ} {\emph {\bibinfo {title}
  {Lectures on the Electrical Properties of Materials}}},\ Oxford science
  publications\ (\bibinfo  {publisher} {Oxford University Press},\ \bibinfo
  {year} {1988})\BibitemShut {NoStop}%
\bibitem [{\citenamefont {Pan}\ and\ \citenamefont {Dou}(2004)}]{Pan2004}%
  \BibitemOpen
  \bibfield  {author} {\bibinfo {author} {\bibfnamefont {A.~V.}\ \bibnamefont
  {Pan}}\ and\ \bibinfo {author} {\bibfnamefont {S.}~\bibnamefont {Dou}},\
  }\href {\doibase https://doi.org/10.1063/1.1763224} {\bibfield  {journal}
  {\bibinfo  {journal} {Journal of Applied Physics}\ }\textbf {\bibinfo
  {volume} {96}},\ \bibinfo {pages} {1146} (\bibinfo {year}
  {2004})}\BibitemShut {NoStop}%
\bibitem [{\citenamefont {Graef}(2001)}]{Miai_2001}%
  \BibitemOpen
  \bibfield  {author} {\bibinfo {author} {\bibfnamefont {M.~D.}\ \bibnamefont
  {Graef}},\ }\href@noop {} {\emph {\bibinfo {title} {Magnetic imaging and its
  applications to materials}}},\ \bibinfo {series} {Experimental methods in the
  physical sciences}, Vol.~\bibinfo {volume} {36}\ (\bibinfo  {publisher}
  {Academic Press},\ \bibinfo {address} {San Diego ; London},\ \bibinfo {year}
  {2001})\BibitemShut {NoStop}%
\bibitem [{\citenamefont {Budker}\ and\ \citenamefont
  {Jackson~Kimball}(2013)}]{Budker_2013}%
  \BibitemOpen
  \bibfield  {author} {\bibinfo {author} {\bibfnamefont {D.}~\bibnamefont
  {Budker}}\ and\ \bibinfo {author} {\bibfnamefont {D.~F.}\ \bibnamefont
  {Jackson~Kimball}},\ }\href@noop {} {\emph {\bibinfo {title} {Optical
  Magnetometry}}}\ (\bibinfo  {publisher} {Cambridge University Press},\
  \bibinfo {address} {New York},\ \bibinfo {year} {2013})\BibitemShut {NoStop}%
\bibitem [{\citenamefont {Jooss}\ \emph {et~al.}(2002)\citenamefont {Jooss},
  \citenamefont {Albrecht}, \citenamefont {Kuhn}, \citenamefont {Leonhardt},\
  and\ \citenamefont {Kronm\"{u}ller}}]{Jooss2002}%
  \BibitemOpen
  \bibfield  {author} {\bibinfo {author} {\bibfnamefont {C.}~\bibnamefont
  {Jooss}}, \bibinfo {author} {\bibfnamefont {J.}~\bibnamefont {Albrecht}},
  \bibinfo {author} {\bibfnamefont {H.}~\bibnamefont {Kuhn}}, \bibinfo {author}
  {\bibfnamefont {S.}~\bibnamefont {Leonhardt}}, \ and\ \bibinfo {author}
  {\bibfnamefont {H.}~\bibnamefont {Kronm\"{u}ller}},\ }\href {\doibase
  https://doi.org/10.1088/0034-4885/65/5/202} {\bibfield  {journal} {\bibinfo
  {journal} {Reports on Progress in Physics}\ }\textbf {\bibinfo {volume}
  {65}},\ \bibinfo {pages} {651} (\bibinfo {year} {2002})}\BibitemShut
  {NoStop}%
\bibitem [{\citenamefont {Kardjilov}\ \emph {et~al.}(2011)\citenamefont
  {Kardjilov}, \citenamefont {Manke}, \citenamefont {Hilger}, \citenamefont
  {Strobl},\ and\ \citenamefont {Banhart}}]{Kardjilov2011}%
  \BibitemOpen
  \bibfield  {author} {\bibinfo {author} {\bibfnamefont {N.}~\bibnamefont
  {Kardjilov}}, \bibinfo {author} {\bibfnamefont {I.}~\bibnamefont {Manke}},
  \bibinfo {author} {\bibfnamefont {A.}~\bibnamefont {Hilger}}, \bibinfo
  {author} {\bibfnamefont {M.}~\bibnamefont {Strobl}}, \ and\ \bibinfo {author}
  {\bibfnamefont {J.}~\bibnamefont {Banhart}},\ }\href {\doibase
  https://doi.org/10.1016/S1369-7021(11)70139-0} {\bibfield  {journal}
  {\bibinfo  {journal} {Materials Today}\ }\textbf {\bibinfo {volume} {14}},\
  \bibinfo {pages} {248} (\bibinfo {year} {2011})}\BibitemShut {NoStop}%
\bibitem [{\citenamefont {Kardjilov}\ \emph {et~al.}(2018)\citenamefont
  {Kardjilov}, \citenamefont {Hilger}, \citenamefont {Manke}, \citenamefont
  {Strobl},\ and\ \citenamefont {Banhart}}]{Kardjilov2018}%
  \BibitemOpen
  \bibfield  {author} {\bibinfo {author} {\bibfnamefont {N.}~\bibnamefont
  {Kardjilov}}, \bibinfo {author} {\bibfnamefont {A.}~\bibnamefont {Hilger}},
  \bibinfo {author} {\bibfnamefont {I.}~\bibnamefont {Manke}}, \bibinfo
  {author} {\bibfnamefont {M.}~\bibnamefont {Strobl}}, \ and\ \bibinfo {author}
  {\bibfnamefont {J.}~\bibnamefont {Banhart}},\ }\href@noop {} {\bibfield
  {journal} {\bibinfo  {journal} {Journal of imaging}\ }\textbf {\bibinfo
  {volume} {4}},\ \bibinfo {pages} {23} (\bibinfo {year} {2018})}\BibitemShut
  {NoStop}%
\end{thebibliography}%

\end{document}